\documentclass[fleqn,usenatbib]{mnras}

\usepackage{newtxtext,newtxmath}
\usepackage[T1]{fontenc}

\DeclareRobustCommand{\VAN}[3]{#2}
\let\VANthebibliography\thebibliography
\def\thebibliography{\DeclareRobustCommand{\VAN}[3]{##3}\VANthebibliography}

\usepackage{graphicx}	% Including figure files
\usepackage{amsmath}	% Advanced maths commands
\usepackage{soul} 
\usepackage{multirow}
\title[SN~2019hcc: A Type II Supernova Displaying Early {O~\sc{ii}} Lines]{SN~2019hcc: A Type II Supernova Displaying Early {O~\sc{ii}} Lines}

\author[E. Parrag et al.]{
Eleonora Parrag$^{1}$,\thanks{E-mail: parrag@cardiff.ac.uk (EP)}
Cosimo Inserra$^{1}$,
Steve Schulze$^{2}$,
Joseph Anderson$^{3}$,
Ting-Wan Chen$^{2,4}$,
\newauthor
Giorgios Leloudas$^{5}$,
Lluis Galbany$^{6}$,
Claudia P. Guti\'errez$^{7,8}$,
Daichi Hiramatsu$^{9,10}$,
\newauthor
Erkki Kankare$^{8}$,
Tomás E. M$\ddot{\rm u}$ller-Bravo$^{11}$,
Matt Nicholl$^{12}$,
Giuliano Pignata$^{13,14}$,
\newauthor
Regis Cartier$^{15}$,
Mariusz Gromadzki$^{16}$,
Alexandra Kozyreva$^{17}$,
Arne Rau$^{4}$,
\newauthor
Jamison Burke$^{9,10}$,
D. Andrew Howell$^{9,10}$,
Curtis McCully$^{9}$,
and
Craig Pellegrino$^{9,10}$
\\
$^{1}$School of Physics \& Astronomy, Cardiff University, Queens Buildings, The Parade, Cardiff, CF24 3AA, UK\\
$^{2}$The Oskar Klein Centre, Department of Astronomy, Stockholm University, AlbaNova, SE-10691 Stockholm, Sweden\\
$^{3}$European Southern Observatory, Alonso de Cordova 3107, Casilla 19, Santiago, Chile\\
$^{4}$Max-Planck-Institut f{\"u}r Extraterrestrische Physik, Giessenbachstra\ss e 1, 85748, Garching, Germany\\
$^{5}$DTU Space, National Space Institute, Technical University of Denmark, Elektrovej 327, DK-2800 Kgs. Lyngby, Denmark \\
$^{6}$Departamento de F\'isica Te\'orica y del Cosmos, Universidad de Granada, E-18071 Granada, Spain.\\
$^{7}$Finnish Centre for Astronomy with ESO (FINCA), FI-20014 University of Turku, Finland\\
$^{8}$Tuorla Observatory, Department of Physics and Astronomy, FI-20014 University of Turku, Finland\\
$^{9}$Las Cumbres Observatory, 6740 Cortona Drive, Suite 102, Goleta, CA 93117-5575, USA\\
$^{10}$Department of Physics, University of California, Santa Barbara, CA 93106-9530, USA\\
$^{11}$School of Physics and Astronomy, University of Southampton, Southampton, Hampshire, SO17 1BJ, UK\\
$^{12}$Birmingham Institute for Gravitational Wave Astronomy and School of Physics and Astronomy, University of Birmingham, Birmingham B15 2TT, UK\\
$^{13}$Departamento de Ciencias Fisicas, Universidad Andres Bello, Avda. Republica 252, Santiago, Chile\\
$^{14}$Millennium Institute of Astrophysics (MAS), Nuncio Monseñor Sotero Sanz 100, Providencia, Santiago, Chile\\
$^{15}$Cerro Tololo Inter-American Observatory, NSF's National Optical-Infrared Astronomy Research Laboratory, Casilla 603, La Serena, Chile\\
$^{16}$Astronomical Observatory, University of Warsaw, Al. Ujazdowskie 4, 00-478 Warszawa, Poland \\
$^{17}$ Max-Planck-Institut f\"ur Astrophysik, Karl-Schwarzschild-Str. 1, D-85748, Garching, Germany \\
}

\date{Accepted XXX. Received YYY; in original form ZZZ}

\pubyear{2020}

\DeclareUnicodeCharacter{2248}{-}
\UseRawInputEncoding
\begin{document}
\label{firstpage}
\pagerange{\pageref{firstpage}--\pageref{lastpage}}
\maketitle

% Abstract of the paper
\begin{abstract}
We present optical spectroscopy together with ultraviolet, optical and near-infrared photometry of SN~2019hcc, which resides in a host galaxy at redshift 0.044, displaying a sub-solar metallicity. The supernova spectrum near peak epoch shows a `w' shape at around 4000~$\Angstrom$ which is usually associated with {O~\sc{ii}} lines and is typical of Type I superluminous supernovae. %Similar features can be produced in this wavelength region however it is concluded these are not what is seen in SN~2019hcc.  
SN~2019hcc post-peak spectra show a well-developed H$\alpha$ P-Cygni profile from 19 days past maximum and its light curve, in terms of its absolute peak luminosity and evolution, resembles that of a fast-declining Hydrogen-rich supernova (SN IIL). The object does not show any unambiguous sign of interaction as there is no evidence of narrow lines in the spectra or undulations in the light curve. Our {\sc tardis} spectral modelling of the first spectrum shows that Carbon, Nitrogen and Oxygen (CNO) at 19000~K reproduce the `w' shape and suggests that a combination of non-thermally excited CNO and metal lines at 8000~K could reproduce the feature seen at 4000~$\Angstrom$. The Bolometric light curve modelling reveals that SN~2019hcc could be fit with a magnetar model, showing a relatively strong magnetic field (B$>3\times10^{14}$G), which matches the peak luminosity and rise time without powering up the light curve to superluminous luminosities. The high-energy photons produced by the magnetar would then be responsible for the detected {O~\sc{ii}} lines. % The overall evolution of SN~2019hcc suggests that magnetars could be the remnant of supernovae at a normal luminosity range with minimal alterations to their evolution. 
As a consequence, SN~2019hcc shows that a `w' shape profile at around 4000~$\Angstrom$, usually attributed to {O~\sc{ii}}, is not only shown in superluminous supernovae and hence it should not be treated as the sole evidence of the belonging to such a supernova type.
\end{abstract}

% Select between one and six entries from the list of approved~Keywords.
% Don't make up new ones.

%%%%%%%%%%%%%%%%%%%%%%%%%%%%%%%%%%%%%%%%%%%%%%%%%%

%%%%%%%%%%%%%%%%% BODY OF PAPER %%%%%%%%%%%%%%%%%%

\section{Introduction}

%\textcolor{red}{Writing the paper pay attention to the citations and use \citep{} and \citet{} rather than \cite only as they are more flexible. See following examples:\\
%\citep{Extreme_SN}\\
%\citet{Extreme_SN}\\
%\citep[bla1,][bla2]{Extreme_SN}\\
%\citet[bla1][bla2]{Extreme_SN}
%}

Historically, supernovae (SNe) were initially classified according to specific observational characteristics, and then a physically motivated classification scheme was built, providing insight into explosion physics and stellar evolution pathways. SNe can be broadly classified into two main types - those which show Hydrogen lines (Type II) and those which do not (Type I). Core-collapse of a massive star with a retained Hydrogen envelope produces the Hydrogen-rich Type II SNe, whereas if such envelope has been stripped we observe stripped envelope supernovae (SESNe), which fall into the Hydrogen-poor Type I. %, as well as thermonuclear SNe. 

SNe~II are considered a single population \citep{Minkowski_1941} but a large spectral and photometric diversity is nowadays observed \cite[e.g.][]{Gutierrez_2017a}. SNe~II were historically split into two categories based on their photometric evolution, SNe~IIL showing a linear decline in the light curve \citep{Barbon_1979} and SNe~IIP showing a plateau for several weeks. \cite{SNIIbook} suggested that the difference in Type IIL, a typically brighter subclass of Type II supernovae, could be due to the presence of a magnetar. However, \cite{Anderson_2014b} suggested that the diversity observed in SN~II light-curves and their spectra is due to the mass and density profile of the retained Hydrogen envelopes. For years, it has been a matter of dispute whether IIL and IIP are a continuous population or have distinctly different physics and progenitors but, recently, increasing evidence has suggested that they are coming from a continuous populations \citep[e.g.][]{Anderson_2014b,Sanders_2015,Galbany_2016,Valenti_2016,deJaeger_2018}. \cite{Anderson_2014b} also noted that very few SNe~II actually fit the classical description of SNe~IIL as most show a plateau of some form. However, \cite{Davis_2019} performed a spectroscopic analysis in the near-infrared (NIR) which found distinct populations corresponding to fast (SN~IIL) and slow (SN~IIP) decliners, though they suggested this could alternatively be accounted for by a gap in the data set.

Further splittings of SNe~II are based on spectroscopic features. SNe~IIb are transitional events between Hydrogen-rich SNe~II and Hydrogen-poor SNe~Ib \citep[e.g.][]{Filippenko_1993}. SNe~IIn display narrow emission lines attributed to interaction with dense circumstellar material \citep[e.g.][]{Schlegel_1990}. SN classification can be time dependent, as some objects have been observed to dramatically change their observables over time, ranging on timescales from weeks to years. In recent years, wide-field surveys have revealed a large diversity of unusual transients that include extreme transitional objects \citep{Modjaz_2019}. One such example is SN~2017ens \citep{Chen_2018}, a transition between a luminous broadline SN~Ic and a SN~IIn. SN~2017ivv is another, sharing properties with fast-declining SN~II and SN~IIb \citep{Gutierrez_2020}, or SN 2014C, which underwent a change from a SN~Ib to SN~IIn due to interaction with a Hydrogen-rich CSM \citep{Milisavljevic_2015}.
Objects such as these can support physical continuity between progenitors and explosion mechanisms of different types \citep{Filippenko_1988}.

Another finding of the wide-field survey has been the discovery of a population of ultra-bright `superluminous' supernovae \citep{Quimby_2011}. SLSNe are intrinsically rare with respect to common core-collapse SNe \citep{Quimby_2013,McCrum_2015,Extreme_SN}, with a recent measurement by \cite{Frohmaier_2021} reporting a local ratio of SLSNe~I to all types of CCSNe of $\sim 1/3500^{+2800}_{-720}$. SLSNe are characterised by absolute luminosities at maximum light of approximately -21~mag \citep{GalYam_2012,Extreme_SN}, though recent evidence suggests that SLSNe in fact occupy a wider range of luminosities, with peak luminosities reportedly as faint as -20~mag \citep[e.g.][]{Angus_2019}. They are typically found in dwarf, metal-poor and star-forming galaxies, suggesting that SLSNe are more effectively formed in low metallicity environments \citep[e.g.][]{Lunnan_2014,leloudas_2015,host_galaxy,Schulze_2018}. Type I superluminous supernovae (SLSNe~I) display a lack of H or He features, and early-time spectra show prominent broad absorption features around 4200~$\Angstrom$ and 4400~$\Angstrom$. These are usually associated with {O~\sc{ii}}, consisting of a complex blend of many individual lines \citep{Quimby_2011,GalYam_2019}.

Here we present the data and analysis of SN~2019hcc, which appears to show typical features of both SLSNe~I and SN~II at different stages in its evolution. The first spectrum appeared to contain a `w' shape associated with {O~\sc{ii}} lines near maximum, typical of SLSNe~I \citep[e.g.][]{Quimby_2011,Extreme_SN}. However, subsequent spectra identify SN~2019hcc as a moderately bright Type II supernova, similar to those discussed in \cite{moderates}, due to the presence of Balmer lines. This is the first such object (to our knowledge) to be identified in the literature.

In this paper, we will show that SN~2019hcc, despite displaying a `w' shape profile similar to those observed in SLSNe~I, otherwise conforms with the typical properties of SNe~II. We will then investigate possible mechanisms which could be responsible for producing such a `w' shape profile in a SN~II. This paper is organised as follows. In Section~\ref{Observations} we report the observations and how the data was obtained and reduced. In Section~\ref{Host} the host galaxy and its properties are analysed. In Section~\ref{Photometry} the rise time and explosion epoch are determined, and the photometry is presented. Section~\ref{Light_Curve_Analysis} contains a detailed analysis of the optical, NIR and bolometric light curve properties. Section~\ref{Spectroscopy} focuses on the spectra of SN~2019hcc, their comparison with other SN types which share common features, and on a close analysis of the Balmer profiles to look for signatures of interaction. Section~\ref{OII lines} considers the  `w'  profile, investigating the required conditions for the formation of the features, and discusses the merit of different powering mechanisms. Section~\ref{Conclusion} provides a summary of our work.

%--------------------------------------------------------------------
\section{Observations and Data Reduction}
\label{Observations}

%-------------------------------------- Two column figure (place early!)
SN~2019hcc was discovered by the \textit{Gaia} satellite \citep{Gaia} as Gaia19cdu on MJD~58640 \citep{Gaia_astronote}, and subsequently by the Asteroid Terrestrial-impact Last Alert System \citep[ATLAS;][]{ATLAS,Smith_2020} on MJD~58643 as ATLAS19mgw \citep{SN2019hcc_Astronote}. The first spectrum was taken on MJD~58643, 3 days after discovery and 7 days after the photometric maximum, see Section \ref{Light_Curve_Analysis}. It was then classified on MJD~58643 as a SLSN~I \citep{classification_astronote} as a consequence of the w-shaped absorption feature around 4000~$\Angstrom$. The redshift was found to be z = 0.044 from the host galaxy emission lines as visible from the second spectrum, and then confirmed by the host galaxy spectrum taken at the end of the SN campaign. We assume a flat $\Lambda$CDM universe with a Hubble constant of $H_0$ = 70~km s$^{\rm{-1}}$Mpc$^{\rm{-1}}$ and $\Omega_m$ = 0.3 and hence a luminosity distance of 194.8~Mpc.

%SC7J004890, an unclassified source found in the GSC v2.3 catalogue at B=20.59 mag,

\begin{figure}
\centering
\includegraphics[width=1\linewidth]{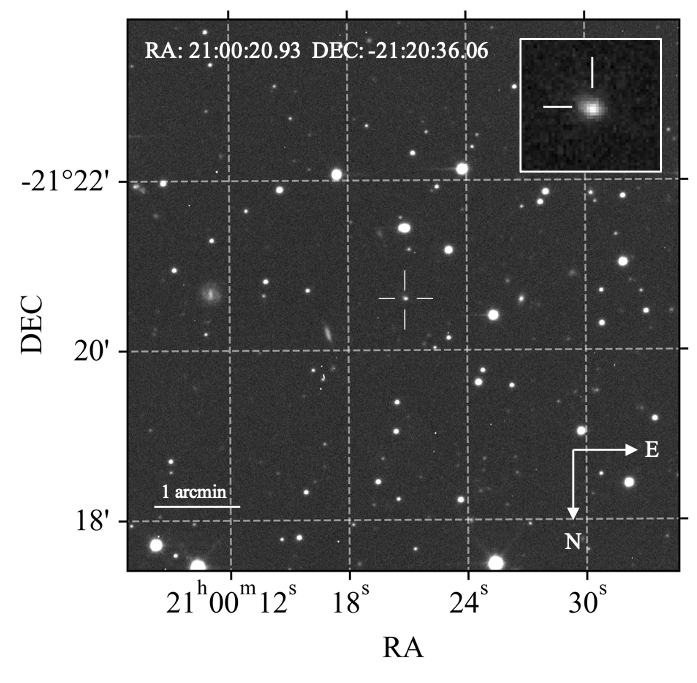}
\vspace{-1mm}
\caption{The finder chart for SN~2019hcc displaying the local environment, taken in $r-$band at MJD~=~58660 by LCO. The host is a low luminosity galaxy. SN~2019hcc is marked by the white crosshairs, and in the blow-up image in the top-right corner.}
\label{fig: finder}
\end{figure}

However, the second spectrum taken on MJD~58655 showed a prominent H$\alpha$ profile implying the target was not a SLSN~I, but rather a bright Type II. It had equatorial coordinates of RA: 21:00:20.930, DEC: -21:20:36.06, with the most likely host J210020.73-212037.2 in the WISEA catalogue at $M_r$ = 19.3 mag \citep{WISEA}, since the redshift of this host and that of SN~2019hcc are matched. %The object's redshift was found to be z = 0.044 based on host emission lines \citep{SN~2019hcc_Astronote}, and 
The Milky Way extinction was taken from the all-sky Galactic dust-extinction survey \citep{Schlafly_2011} as Av~=~0.19. Taking Rv = 3.1, this gives an E(B-V) of 0.06. Since there are no {Na~\sc{i}} D absorption lines related to the host and the SN luminosity and colour evolution appear to be as expected in a SN~II (see Section \ref{Light_Curve_Analysis}), the host galaxy reddening has been assumed negligible.
Figure \ref{fig: finder} shows the finder chart and the local environment of SN~2019hcc.

\subsection{Data Reduction}
Five optical spectra were taken over a range of 5 months with the NTT+EFOSC2 at the La Silla Observatory, Chile. This was under the advanced Public ESO Spectroscopic Survey of Transient Objects programme \citep[ePESSTO+;][]{Smartt_2015}. This was alongside a host galaxy spectrum taken over a year after explosion when SN~2019hcc was no longer visible. The spectra were reduced using the PESSTO NTT pipeline\footnote{https://github.com/svalenti/pessto}. %This is a package to reduce EFOSC and SOFI data using standard methods \citep{Smartt_2015}. 
There was also one spectrum taken by the Goodman High Throughput Spectrograph at the Southern Astrophysical Research telescope (SOAR) \citep{Clemens_2004}, reduced using the dedicated pipeline \citep{Sanchez_2019}. %\footnote{http://www.ctio.noao.edu/soar/content/goodman-data-reduction-pipeline}. 

Photometric data was obtained by the Las Cumbres Observatory \citep[LCO;][]{Brown_2011} with the camera Sinistro built for the 1m-class LCO telescopes, and by the Liverpool Telescope \citep[LT;][]{Steele_2004} on the Canary Islands. Images were combined using SNOoPY \footnote{SNOoPy is a package for SN photometry using PSF fitting and/or template subtraction developed by E. Cappellaro. A package description can be found at http://sngroup.oapd.inaf.it/snoopy.html} and the magnitudes were retrieved using PSF photometry, with the zero-point calibration completed using reference stars accessed from the Panoramic Survey Telescope and Rapid Response System \citep[Pan-STARRS;][]{Chambers_2016} and the Vizier catalogues \citep{Vizier}. This was performed using the code described in detail in Appendix \ref{Appendix: Code}. Additional photometry was also taken by ATLAS, \textit{ Swift} + Ultraviolet/Optical Telescope \cite[UVOT;][]{Roming_2005} and the Gamma-Ray Burst Optical and Near-Infrared Detector \citep[GROND;][]{GROND}. GROND is an imaging instrument to investigate Gamma-Ray Burst Afterglows and other transients simultaneously in seven bands $grizJHK$ mounted at the 2.2~m MPG telescope at the ESO La Silla Observatory (Chile). The GROND images of SN~2019hcc were taken under the GREAT survey \citep{Chen_2018}. GROND \citep{Kruhler_2008}, ATLAS and \textit{ Swift} data were reduced using their own pipelines. The photometry and spectroscopy logs, including dates, configurations, and magnitudes are reported in Appendix \ref{Appendix: Data}. As \textit{Swift} observes simultaneously with UVOT and the X-ray Telescope (XRT), we report that the corresponding upper limit on the unabsorbed 0.3-10~keV flux is 2.6x10$^{-14}$~cgs
(assuming a power law with photon index 2 and the Galactic column density of 4.9x10$^{20}$~cm$^{-2}$) resulting in an upper limit on luminosity of $\sim$10$^{41}$~erg/s at SN~2019hcc distance. %(at 187.7~Mpc).
The closest non-detections were taken by ATLAS from 34~days to 22~days before discovery, with a confidence of 3$\sigma$.

\section{Host Galaxy}
\label{Host}
%{\bf We retrieved SN~2019hcc host galaxy properties with MAGPHYS \citep[Multi-wavelength Analysis of Galaxy Physical Properties,]{daCunha_2008}. This tool provides a physically motivated model to interpret observed spectral energy distributions (SEDs) of galaxies in terms of galaxy-wide physical parameters related to the stars and the interstellar medium. %It works with a library of model SEDs at the same redshift and in the same photometric bands as the observed galaxy, and returns physical parameters through the comparison of the observed SED with all the models in the library. It was used to create a best-fit SED for the galaxy, using the magnitudes and corresponding errors measured from the host images (using code described in Appendix \ref{Appendix: Code}) with the $griz-$bands as the input. 
%The host SED has been retrieved via images taken at the Liverpool Telescope \citep[LT]{Steele_2004}, roughly 1 year after first detection when the SN was not longer visible. Figure \ref{fig: magphys} shows the best-fit SED. }

%\cite{Zhou_2020} analysed a sample of 29 Core-collapse supernova (CCSN) host galaxies, and they found a median log(M*/M$_\odot$) of 10.26$\pm$0.56 and a star formation rate (SFR) M$_\odot$ yr$^{-1}$ of 0.14$\pm$0.76 - this suggests that the host of SN~2019hcc has an unusually low mass, with the SFR of the host within the error of this sample value.

The host galaxy spectrum for SN~2019hcc was taken with NTT+EFOSC2 \citep{Buzzoni_1984} at the La Silla Observatory, Chile, on MJD~59149, when the SN was no longer visible, as part of the ePESSTO+ programme \citep{Smartt_2015}. The line fluxes were measured using the splot function in \textsc{iraf} \citep{IRAF} by taking a number of measurements and averaging to account for the uncertainty in the location of the continuum.
The host galaxy spectrum was analysed using pyMCZ.  This is an open-source Python code which determines the metallicity indicator, Oxygen abundance (12 + log(O/H)), through Monte Carlo sampling, and gives a statistical confidence region \citep{Bianco_2016}. The input of this code is the line flux and associated uncertainties for lines such as [{O~\sc{ii}}] and H$\alpha$ from the host galaxy spectrum. \cite{Kewley_2008} found that the choice of metallicity calibration has a significant effect on the shape and y-intercept (12+log(O/H)) of the mass-metallicity relation, therefore multiple markers are used to measure the metallicity in an effort to give a representative range.

\begin{figure}
\centering
\includegraphics[width=1\linewidth]{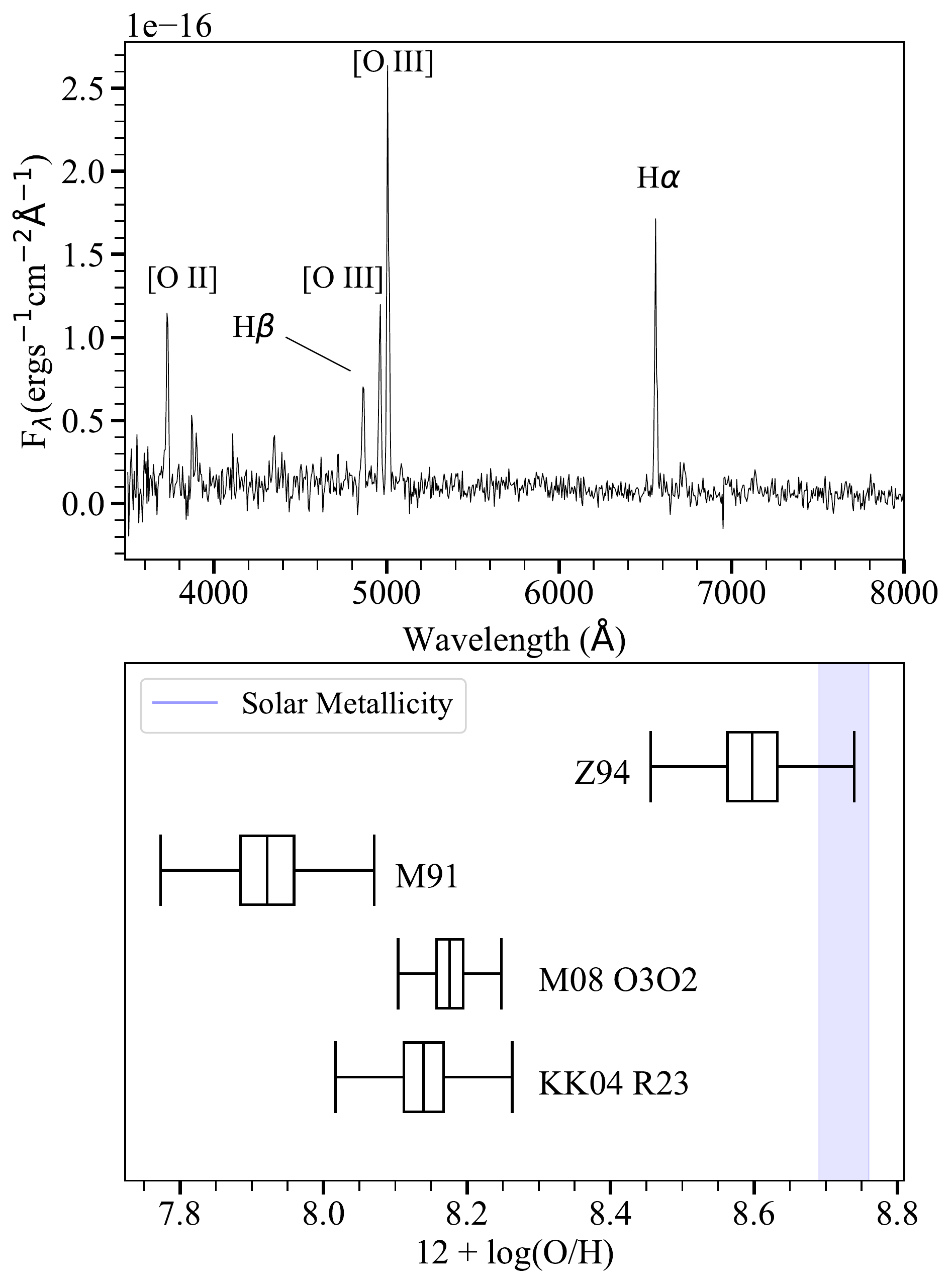}
\vspace{-1mm}
\caption{Top panel: NTT galaxy spectrum used as input for pyMCZ, with the relevant lines labelled. The H$\alpha$/H$\beta$ ratio is $2.2\pm0.1$. The wavelength is in the rest frame. Bottom panel: Reproduced output of pyMCZ, the metallicity measured via several markers is displayed as box plots. The central value is the median (or 50$^{\rm{th}}$ percentile). The inner box represents the inter-quartile range (IQR) - 50$^{\rm{th}}$ to 16$^{\rm{th}}$ percentile and 84$^{\rm{th}}$ to 50$^{\rm{th}}$ percentile (the 16$\%$ is an analogy to the Gaussian 1$\sigma$ interval), whilst the outer bars represent the minimum and maximum data values, excluding outliers. The outliers are those values further than 1.5xIQR from the edges of the IQR. The blue band is a range of solar metallicity values found in literature  - from 8.69 in \protect\cite{Asplund_2009} to 8.76 in \protect\cite{Caffau_2011}.}
\label{fig: gal_spectrum}
\end{figure}

Figure \ref{fig: gal_spectrum} shows the input (upper panel) and output (lower panel) for pyMCZ \citep[see][]{Bianco_2016}. The metallicity estimators are those of \citet[][]{zaritsky_1994} [Z94], \citet[][]{McGaugh_1991} [M91], \citet[][]{Maiolino_2008} [M08], and  \citet[][]{Kewley_2008} [KK04]. These metallicity markers are all based on $R_{23}$, see \cite{Bianco_2016} for a summary and further details:

\begin{equation}
R_{23} = \frac{[\rm{O~II}]\lambda3727 + [\rm{O~III}]\lambda\lambda4959, 5007}{H\beta}
\label{eqn R23}
\end{equation}

[{N~\sc{ii}}] $\lambda$6584 is not visible in this spectrum, and at this resolution it would be very difficult to resolve as it is so close to H$\alpha$. A lack of [{N~\sc{ii}}] is an indicator of low metallicity, therefore the lower branches of the metallicity indicators were used in the code, apart from Z94, %([{O~\sc{ii}}] $\lambda$3727 + [{O~\sc{iii}}] $\lambda\lambda$4959,5007)/H$\beta$ \citep{zaritsky_1994}, 
where only the upper branch is available in pyMCZ. The metallicity markers used are those available given the line fluxes which were input into pyMCZ, which are labelled in the top panel of Figure \ref{fig: gal_spectrum}. Averaging them we obtain a host galaxy metallicity of 12~+~log(O/H)~$= 8.08\pm0.05$, which is below solar abundance.

We also note that the H$\alpha$/H$\beta$ flux ratio in the host spectrum is measured to be 2.2~$\pm$~0.1, less than the intrinsic ratio 2.85 for case B recombination at T~=~10$^4$~K and n$_e$ $\sim10^2-10^4$~cm$^{-3}$ \citep{osterbrock_1989}. A ratio of less than 2.85 can result from an intrinsically low reddening combined with errors in the stellar absorption correction and/or errors in the line flux calibration and measurement \citep{Kewley_2008}.

%\cite{Galbany_2018} found an average Type II host mass of 10.11 dex, a star formation rate density of -1.809$\pm$0.119
%\cite{Lunnan_2014} {\bf , low stellar mass (2 × 10$^_{\rm 8}$~M$_\odot$) population, with a high median specific star formation rate (sSFR ≈ 2~Gyr$^{\rm -1}$).}

Models by \cite{Dessart_2014} hint to a lack of SNe~II below 0.4~Z$_{\odot}$. However, this may be biased as higher luminosity hosts were used which tend to have higher metallicity. On the other hand, SLSNe~I are predominantly found in dwarf galaxies, indicating that their progenitors have a low metallicity. A 0.5~$Z_{\odot}$ threshold has been suggested for the formation of SLSNe~I \citep{host_galaxy}. \cite{Lunnan_2014} found a median metallicity of 8.35 = 0.45~Z$_{\odot}$ for a sample of 31 SLSNe~I. 

The measured metallicity was compared to both Type II and SLSN~I hosts. Table \ref{tab: gal_prop} contains the mean metallicity excluding Z94 (this is likely incorrect as it is the upper branch) from Figure \ref{fig: gal_spectrum}, compared to averages for SLSNe~I and SNe~II. \cite{Schulze_2020} performed a comprehensive analysis of SN hosts based on a sample of 888 SNe of 12 distinct classes, and found a median metallicity 12~+~log(O/H)~$= 8.26^{+0.26}_{-0.30}$ for a sample of 37 SLSNe~I. \cite{Galbany_2018} presented a compilation of 232 SN host galaxies, of which 95 were Type II hosts with an average metallicity (12 + log(O/H)) of 8.54~$\pm$~0.04. The mean metallicity for SN~2019hcc is within the range of the SLSN~I host metallicity found by \cite{Schulze_2020}, and is low compared to the average metallicity of Type II hosts.

The host galaxy absolute magnitude was measured to be $-15.8\pm0.3$ in $r-$band and $-15.8\pm0.2$ in $B-$band. \cite{Gutierrez_2018} defined a faint host as having $M_r \gtrsim -18.5$~mag, and analysed the hosts of a sample of low-luminosity SNe~II, finding a mean host luminosity of $-16.42\pm0.39$~mag. \cite{Anderson_2016} examined a sample of SNe~II in a variety of host types and found a mean host luminosity $M_r$ of $-20.26\pm0.14$~mag. For SLSNe~I, \cite{Lunnan_2014} found a low average magnitude ($M_{B} \approx - $17.3 mag). Table \ref{tab: gal_prop} also contains the average $M_B$ magnitudes for both SLSNe~I and SNe~II from \cite{Schulze_2020}. SN~2019hcc has a lower luminosity and metallicity host with respect to the average value for SNe~II and SLSNe~I reported in the literature (see Table~\ref{tab: gal_prop}).

We retrieved further SN~2019hcc host galaxy properties by modelling the spectral energy distribution (SED) using the software package Prospector version 0.3 \citep{Leja2017a, Johnson2019a}. An underlying physical model is generated using the Flexible Stellar Population Synthesis (FSPS) code \citep{Conroy2009a}. A Chabrier initial mass function \citep{Chabrier2003a} is assumed and the star formation history (SFH) is approximated by a linearly increasing SFH at early times followed by an exponential decline at late times (functional form $t \times \exp\left(-t/\tau\right)$). The model was attenuated with the \cite{Calzetti2000a} mode, and a dynamic nested sampling package density \citep{Speagle2020a} was used to sample the posterior probability function. To interface with FSPS in python, python-fsps \citep{ForemanMackey2014a} was used.

The photometry images were sourced from the Panoramic Survey Telescope and Rapid Response System (Pan-STARRS, PS1) Data Release 1 \citep{Chambers_2016}, the \textit{Galaxy Evolution Explorer} (GALEX) general release 6/7 \citep{Martin2005a}, the ESO VISTA Hemisphere Survey \citep{McMahon2013a}, and preprocessed WISE images \citep{Wright2010a} from the unWISE archive \citep{Lang2014a}\footnote{\href{http://unwise.me}{http://unwise.me}}. The unWISE images are based on the public WISE data and include images from the ongoing NEOWISE-Reactivation mission R3 \citep{Mainzer2014a, Meisner2017a}. The host brightness was measured using LAMBDAR\footnote{\href{https://github.com/AngusWright/LAMBDAR}{https://github.com/AngusWright/LAMBDAR}} \citep[Lambda Adaptive Multi-Band Deblending Algorithm in R;][]{Wright2016a} and the methods described in \cite{Schulze_2020}.

\begin{figure}
\centering
\includegraphics[width=1\linewidth]{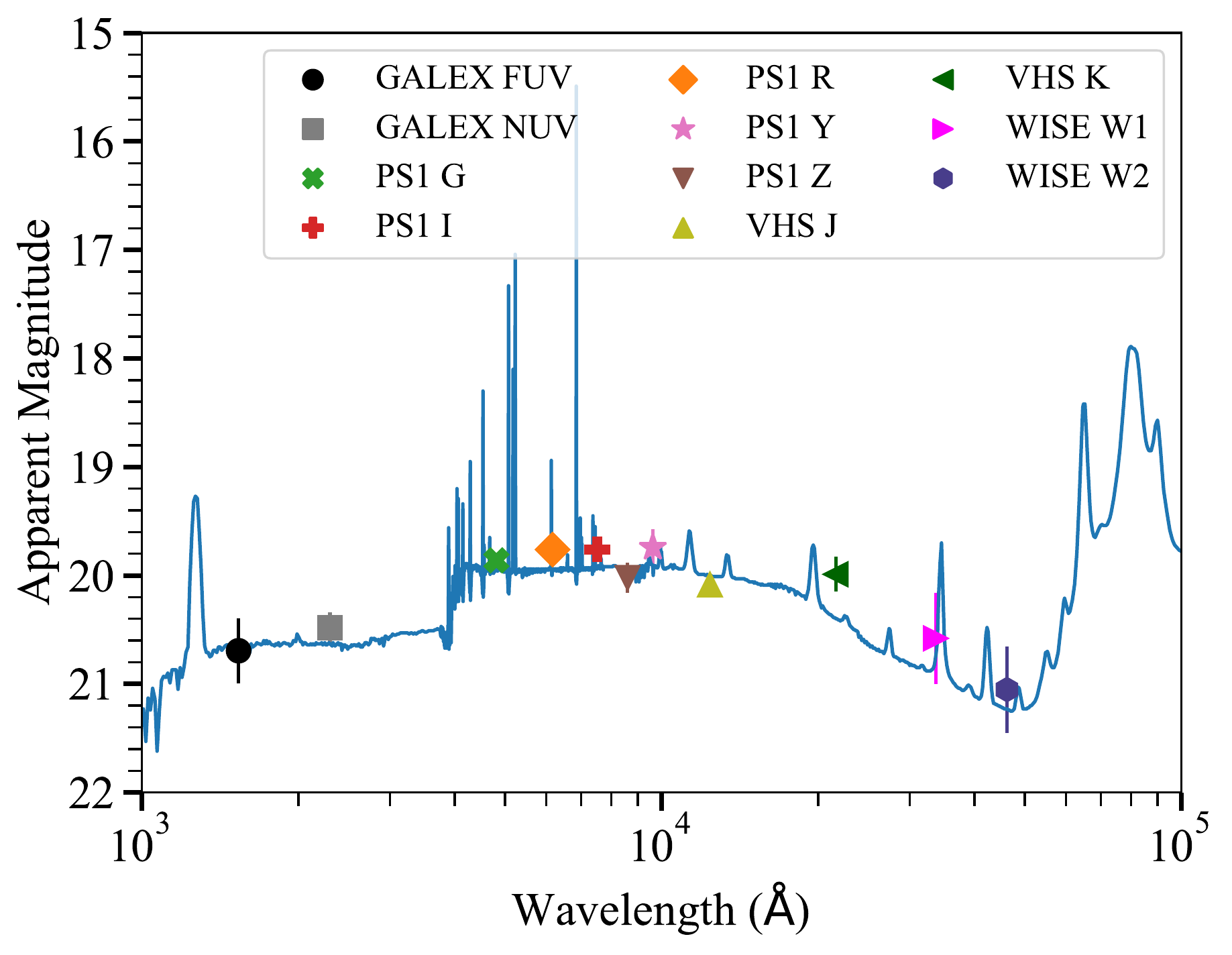}
\vspace{-1mm}
\caption{Galaxy photometry of SN~2019hcc from GALEX, PS1, VHS and WISE, with the best fit SED from Prospector, The median $\chi^{2}$ divided by the number of filters (n.o.f.) is 10.65/11 and includes emission lines from {H~\sc{ii}} regions in the fitting.}
\label{fig: prospector}
\end{figure}

Figure \ref{fig: prospector} shows the best fit SED to the SN~2019hcc photometry for filters GALEX $FUV$ (20.69$\pm$0.30~mag) and $NUV$ (20.48$\pm$0.14~mag), PS1 $GIRYZ$ (19.86$\pm$0.03~mag, 19.76$\pm$0.04~mag, 19.76$\pm$0.04~mag, 19.74$\pm$0.17~mag, 20.02$\pm$0.14~mag), VHS $JK$ (20.08$\pm$0.08~mag, 19.99$\pm$0.16~mag) and WISE $W1$ (20.58$\pm$0.42~mag) and $W2$ (21.05$\pm$0.40~mag). The magnitudes are in the AB system and corrected for Milky Way extinction. Table \ref{tab: gal_prop} shows the galaxies properties inferred from the best-fit SED to the host galaxy photometry. The E(B-V) inferred for SN~2019hcc matches well with the E(B-V) based on the Milky Way extinction. The mass of the host best matches the median SLSN~I host mass, whilst the SFR is low for both SLNSe~I and SNe~II. The age of the SN~2019hcc host has a large uncertainty that covers the range of both classes, and the magnitude is low for both classes. The SFR is significantly lower for SN~2019hcc. However, the mass of the host is lower than the median for both SLSNe~I and SNe~II, and therefore the sSFR falls between the two.

%{\bf To compare the metallicity outputs from MAGPHYS and pyMCZ, the mass fraction of heavy elements Z needs to be converted to the oxygen abundance 12+log(O/H) - this can be done in a simplified way \citep{Chruslinska_2019}: log(Z/Z$_{\odot}$) = Z$_{O/H}$ - Z$_{O/H\odot}$. Assuming a solar abundance of 8.69 \citep{Asplund_2009}, the MAGPHYS value Z/Z$_{\odot}$ = 0.77 would approximately correspond to 12+log(O/H)=~8.58. This is higher than all the metallicity markers except Z94 and is likely an overestimate given that the galaxy spectrum suggests low metallicity.}

\begin{table}
\begin{tabular}{ c c c c }
\hline 
Property & SN~2019hcc & SLSN~I & SN~II  \\ 
\hline 
log(M/M$_{\odot}$) & $7.95^{+0.10}_{-0.33}$ &  $8.15^{+0.23}_{-0.24}$ & 9.65$\pm$0.05\\
SFR (M$_{\odot}$yr$^{-1}$) & $0.07^{+0.04}_{-0.01}$ & $0.59^{+0.22}_{-0.20}$  & 0.58$\pm$0.05 \\
log(sSFR) (yr$^{-1}$) & $-9.10^{+1.42}_{-1.78}$ & $-8.34^{+0.30}_{-0.32}$  & $-9.86\pm$0.02 \\
Age (Myr) & $2971^{+2079}_{-2131}$ & $427^{+119}_{-124}$ & 4074$\pm$188\\
E(B-V) & $0.04^{+0.06}_{-0.03}$ & $0.31^{+0.05}_{-0.04}$ &  0.14$\pm$0.01\\
12+log(O/H) & 8.08$\pm$0.05 & $8.26^{+0.26}_{-0.30}$ & 8.54$\pm$0.04 \\
$M_B$ (mag) & $-15.80\pm0.20$ & $-17.51^{+0.30}_{-0.28}$ & $-19.15\pm0.09$ \\
\hline 
\end{tabular}
\caption{Galaxy properties from \textsc{Prospector} for SN~2019hcc, and median values from \protect\cite{Schulze_2020} for SLSNe~I and SNe~II, excluding 12+log(O/H) for Type II which is from \protect\cite{Galbany_2018}.}
\label{tab: gal_prop}
\end{table}

\begin{figure*}
\centering
\includegraphics[width=1\linewidth]{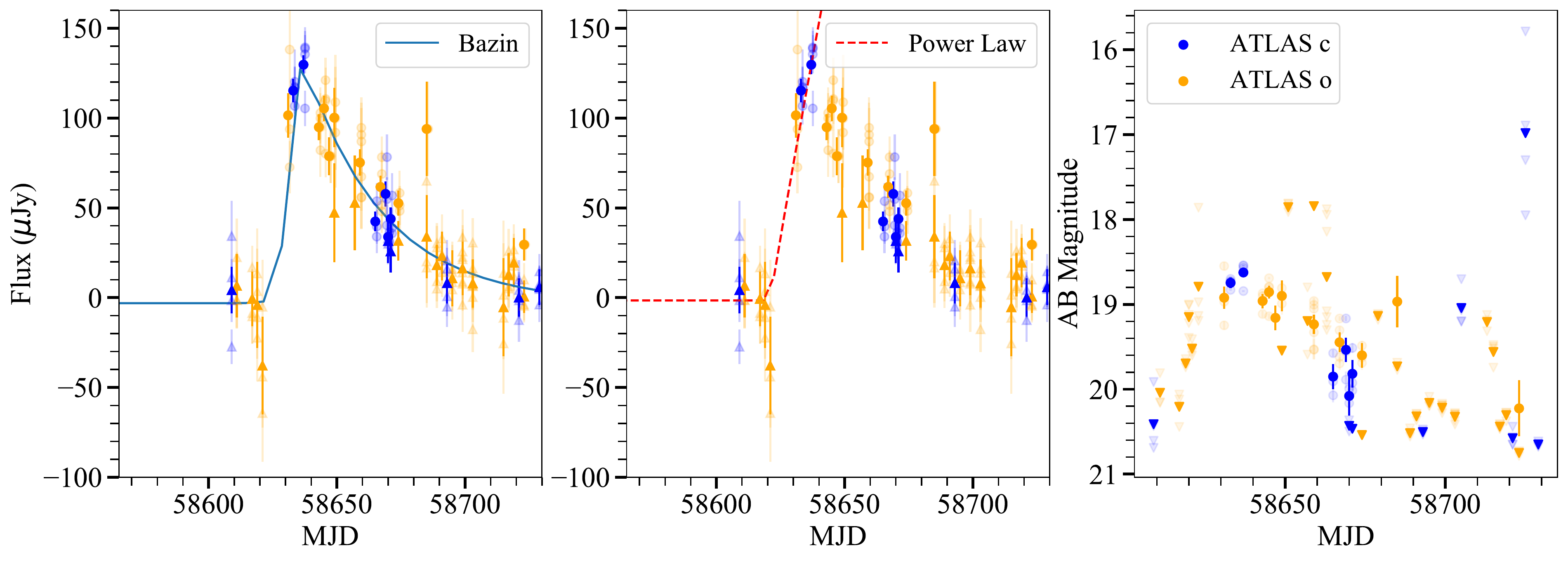}
\vspace{-1mm}
\caption{Left panel: fit to the ATLAS forced photometry weighted mean flux \protect\citep{Bazin_2009}, in order to determine the peak epoch. This finds the maximum epoch to be MJD~$58636.2\pm2.2$.  Points with errors $>30 \mu$Jy have been removed for clarity. Middle panel: a power law fit to the pre-peak flux data (including the upper limits) -  this finds an explosion epoch of MJD~$58621.0\pm7.2$. Upper limits are marked as triangles. Where multiple points from the same epoch were taken, these were averaged - the original points are marked with a lighter hue. Right panel: ATLAS forced photometry weighted mean flux converted to AB magnitude, for orange and cyan filters. Images with flux significance $<3\sigma$ were converted to upper limits.}
\label{fig: BAZIN}
\end{figure*}

%%%%%%%%%%%%%%%%%%%%%%%%%%%%%%%%%%%%%%%%%%%%%%%%%%%%%%%%%%%%%%%%%%%%%%%%%%%%%%%%
\section{Photometry}
\label{Photometry}

\subsection{Rise time and Explosion Epoch}

We determined the rise time and explosion epoch following the methodology presented in \cite{Gonz_Gait_2015}. We applied this approach to the ATLAS data only, both orange and cyan, as it is the only photometry available which covers the pre-peak light curve albeit with many upper limits. It is not ideal to combine different bands however as there are few points it is an unavoidable uncertainty. We then measure the explosion epoch using a power law fit (Equation \ref{eqn 1}) from the earliest pre-peak upper limit to maximum luminosity:

\begin{equation}
\begin{split}
& f(t) = {a(t-t_{\rm exp})}^n & \mbox{ if $t$ > $t_{\rm exp}$} \\
& f(t) = 0 & \mbox{ if $t$ < $t_{\rm exp}$}\,.
\end{split}
\label{eqn 1}
\end{equation}

Here $a$ is a constant and $n$ is the power index, both of which are free parameters, and $t_{\rm exp}$ is the explosion date in days. This fit was done using a least squares fit as implemented by \textsc{scipy.curve\_fit} in Python to the pre-maximum light curve in flux, and the explosion epoch was measured to be MJD~$58621.0\pm7.2$. An alternative method of measuring the explosion epoch is to take the midpoint between the first non-detection and the first detection - this would be between MJD~58609 and MJD~58631, giving an estimate of the explosion epoch of MJD~58620, which is within the errors and consistent with the previous measurement. 

For the epoch of maximum light, we used the phenomenological equation for light curves from \cite{Bazin_2009}. This form, as shown in Equation \ref{eqn 2}, has no physical motivation but rather is flexible enough to fit the shape of the majority of supernova light curves. 

\begin{equation}
f(t) = A \frac{e^{-(t-t_{\rm 0})/t_{\rm fall}}}{1+e^{\rm (t-t_{\rm 0})/t_{\rm rise}}} + B \,.
\label{eqn 2}
\end{equation}

Here $t_{\rm 0}$, $t_{\rm rise}$, $t_{\rm fall}$, $A$ and $B$ are free parameters. The derivative, as seen in Equation \ref{eqn 3}, was used to get the maximum epoch ($t_{\rm max}$), and the uncertainties from the fit were propagated through the below equation \citep{Gonz_Gait_2015}:

\begin{equation}
t_{\rm max} = t_{\rm 0} + t_{\rm rise} \times log(\frac{-t_{\rm rise}}{t_{\rm rise}+t_{\rm fall}})\,.
\label{eqn 3}
\end{equation}

The maximum epoch was found from the Bazin fit to be MJD~$58636.2\pm2.2$ - this was done by fitting to the flux data, see the right panel on Figure \ref{fig: BAZIN}. This will be the maximum hereafter referred to in the paper, and can be approximated as the peak in ATLAS $o-$band, as this is the band the majority of these points are in. Points with an error greater than 30~$\mu$Jy have been removed for clarity. Combining this result with the explosion epoch gives a rise time of $15.2\pm7.5$~days. 

ATLAS $o-$band is close to $R-$band. The average $R-$band rise from the `gold' samples (consisting of 48 and 38 SNe each from different surveys) of SNe~II from \cite{Gonz_Gait_2015} was $14.0^{+19.4}_{-9.8}$~days. %This is a very large error but consistent with the rise found for SN~2019hcc.
\cite{Pessi_2019} reported an average $r-$band rise time for a sample of 73 SNe~II of $16.0\pm3.6$~days. Both results are consistent with our measured value - therefore it seems the rise of SN~2019hcc is typical for a SN~II. In contrast, SLSNe~I light-curves have longer timescales with an average rise of 28 and 52~days for SLSNe~I Fast and Slow respectively \citep{Nicholl_2015,Extreme_SN}. %For the case of magnetar-powered SNe, \cite{magnetar} successfully fitted magnetar models to bright SNe~II SN~2004em and OGLE14\textendash073, which both had rise times on the order of 100~days. With the rise time for SN~2019hcc measured to be 15.3$\pm$7.4~days, this is more similar to the average of a Type II rather than that of SLSNe~I. 
Despite the average longer rise of SLSNe~I to SNe~II, it should be noted that the fastest riser SLSNe~I can have some overlap within the errors of the slowest SNe~II values from \cite{Gonz_Gait_2015}.

\begin{figure*}
\centering
\includegraphics[width=1\linewidth]{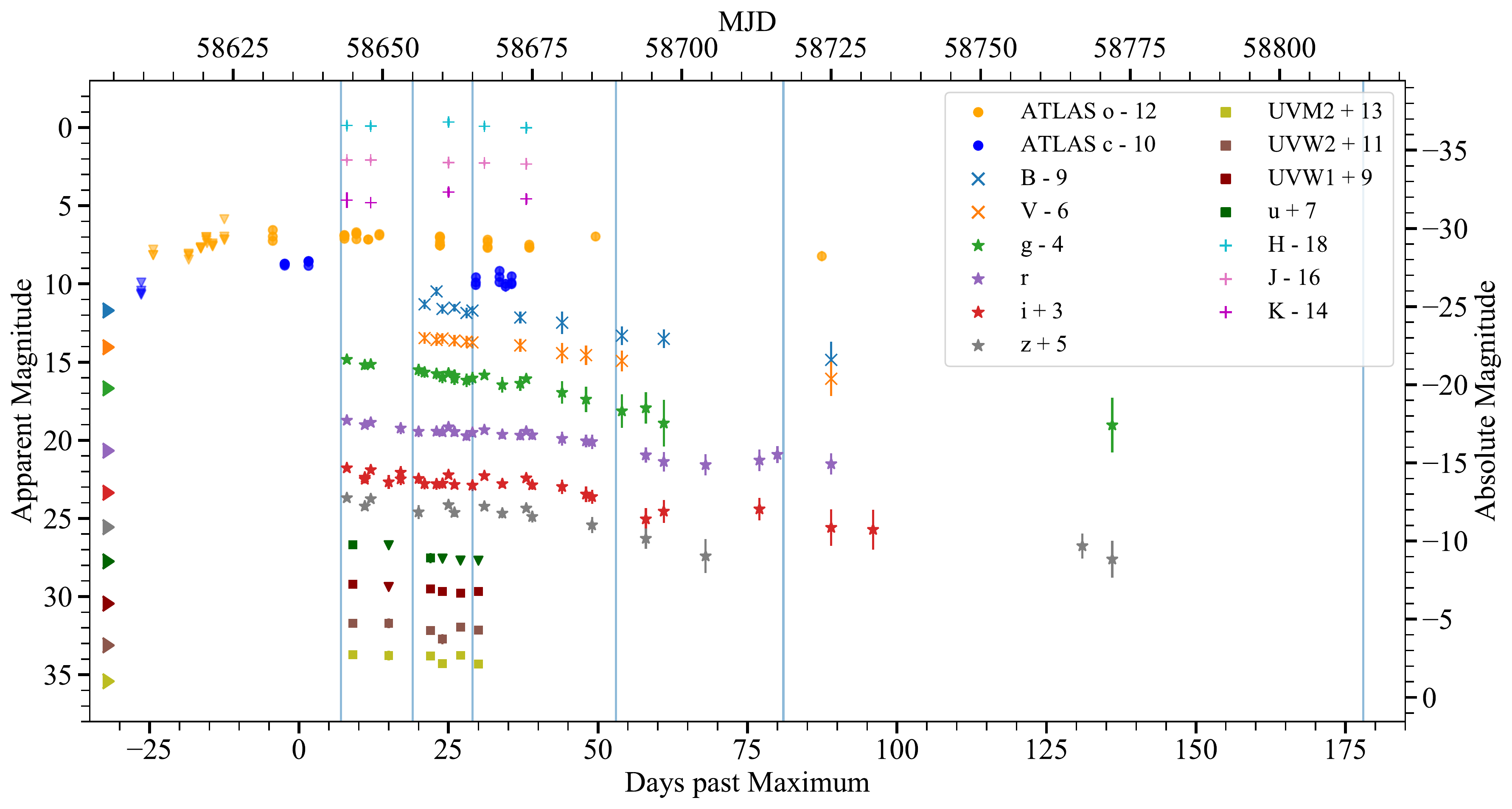}
\vspace{-1mm}
\caption{Photometry for SN~2019hcc  - the light curves from various sources: $BVgri$, bands were taken by LCO, and $griz$ bands were also taken by LT. Alongside this, there is ATLAS data including the pre-peak limits, \textit{ Swift} UV data, and GROND NIR data. The vertical lines mark the epochs when the spectra were taken. The markers on the left y-axis signify the galaxy magnitude in the respective bands.}
\label{fig: Photometry}
\end{figure*}

%%%%%%%%%%%%%%%%%%%%%%%%%%%%%%%%%%%%%%%%%%%%%%%%%%%%%%%%%%%%%%%%%%%%%%%%%%%%%%%%

\subsection{Multi-band light curve}

The majority of photometric data were taken by LCO in bands $BVgri$, and by LT in bands $griz$. The light curve produced from this data was created using a code written using Python packages AstroPy and PhotUtils (see Appendix \ref{Appendix: Code} for further detail). This was complemented by ATLAS data in the orange and cyan bands, UV data from \textit{Swift}, optical ($griz$) and NIR ($JHK$) data from GROND. Figure \ref{fig: Photometry} shows the photometric evolution of SN~2019hcc in all available bands. The UV data covers 21 days, and appears to follow a linear decline. The NIR data covers a similar period of 30 days, and are roughly constant in magnitude. There is a linear decline of $\sim$50 days from peak in all optical bands, with a magnitude change of $\sim$1.5~mag in $r-$band, followed by a steeper drop of $\sim$2~mag from 50 to 70 days. The decline rate is similar in the other bands with the exception of $g-$band which declines faster, at a rate of $\sim$2~mag in the first $\sim$50~day after maximum light, and subsequently $\sim$3~mag in the steeper decline. 
The $BV-$bands data for \textit{Swift} were excluded as they were contaminated by host galaxy light. Such a contamination is far less in $u$, $uvw1$, $uvm2$ and $uvw2$. The \textit{ Swift} detections were at level of 3-4$~\sigma$. GROND $griz$ magnitudes  were not template subtracted as there were no templates available. However, the data were taken soon after maximum light, where the difference between the host galaxy magnitude and that of SN~2019hcc is at its maximum, and therefore should not add significant uncertainty. LT and LCO magnitudes were template subtracted as part of the photometry code described in Appendix \ref{Appendix: Code}.
%this did not make a significant difference, but slightly increases the scatter seen for these bands).

\begin{figure}
\centering
\includegraphics[width=1\linewidth]{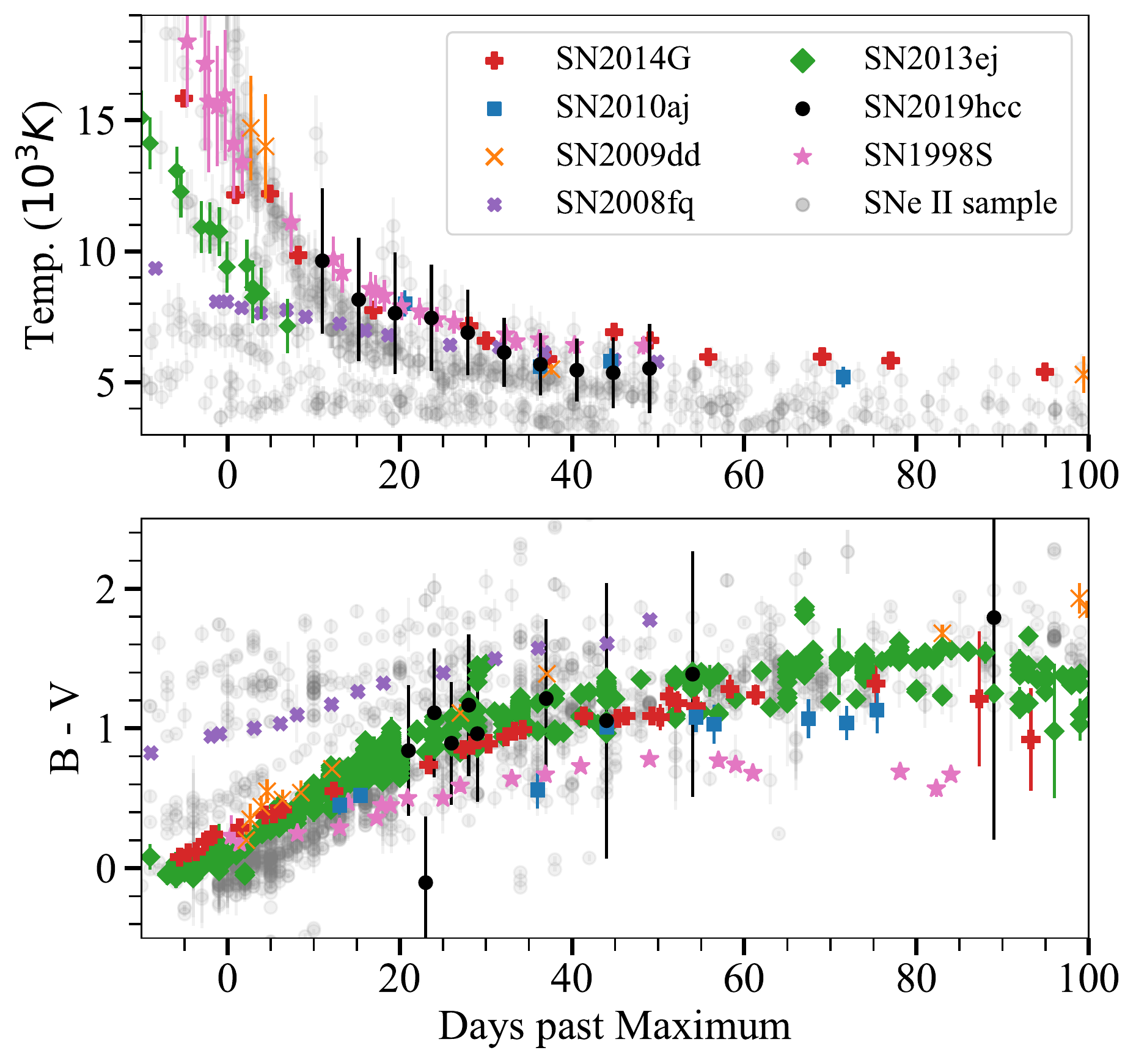}
\vspace{-1mm}
\caption{Top panel: the blackbody temperature evolution - for SN~2019hcc this is from the fit to the photometric data, whilst for the other SNe it is from the literature. The uncertainties for SN~2019hcc are from the curve fit. There were no uncertainties reported in the literature for the temperatures of SN~2014G and SN~2008fq. Bottom panel: colour evolution $B-V$ compared with the same SNe of the upper panel.The temperature and colour evolution from the sample of SNe~II from \protect\cite{Faran_2014} is shown in grey.}
\label{fig: blackbody}
\end{figure}

Figure \ref{fig: blackbody} shows the evolution of the blackbody temperature fit to the photometric data together with the $B-V$ colour  evolution, both for SN~2019hcc and a selection of SNe~II. These are: SN~2013ej \citep{Valenti_2014}, SN~2014G \citep{Terreran_2016}, SN~2008fq \citep{Taddia_2013}, SN~1998S \citep{Fassia_2000,Fassia_2001}, SN~2009dd and SN~2010aj \citep{moderates} together with a sample of 34 SNe~II from \cite{Faran_2014}. SNe~1998S and 2014G - a Type IIn and IIL, respectively - were chosen for their spectroscopic similarity to SN~2019hcc near peak. SN~2013ej, SN~2010aj, and SN~2008fq provide a small sample of well observed SNe~II displaying a similar peak magnitude of SN~2019hcc, which fall in the category of relatively bright Type~II \citep{moderates}. The $griz$ bands for SN~2019hcc were individually interpolated to 10 evenly spaced points across the date range, and the temperature was found by fitting to these bands at each point. The interpolation was done using gaussian processes from \textsc{sklearn}, and the errors from the photometric points were interpolated using interp1d from \textsc{scipy}. For the colour the points were chosen where both $B$ and $V$ were available. The fits for temperature are expected to become worse as the photospheric phase passes and the blackbody approximation is less appropriate. The temperature and colour evolution for the Type II supernovae were taken from the above papers. 
The temperature and colour evolution of the SNe~II sample \citep{Faran_2014} were calculated from the data available on the Open Supernova Catalogue \citep{OSC}. These SNe have a large range within which the temperature evolution falls, and appears to have multiple branches, which spans the range of temperature and colour evolution of the SNe~II selected for a direct comparison. From Figure \ref{fig: blackbody}, it appears the colour and temperature evolution of SN~2019hcc is not unusual with respect to the SNe chosen for a direct comparison or that of \cite{Faran_2014}. Overall, SN~2019hcc colour and temperature evolution appears to closely resemble those of SN~2014G and SN~2009dd. The colour evolution appears to have two regimes, a steeper slope until $\sim$30-40 days followed by a less steep rise. The first slope is 2.8~mag per 100 days which is very similar to the average 2.81~mag per 100 days obtained by \cite{deJaeger_2018} for $B-V$. They also found a transition between the two regimes at 37.7~days which is roughly consistent with what is seen in the colour evolution.

\begin{figure}
\centering
\includegraphics[width=1\linewidth]{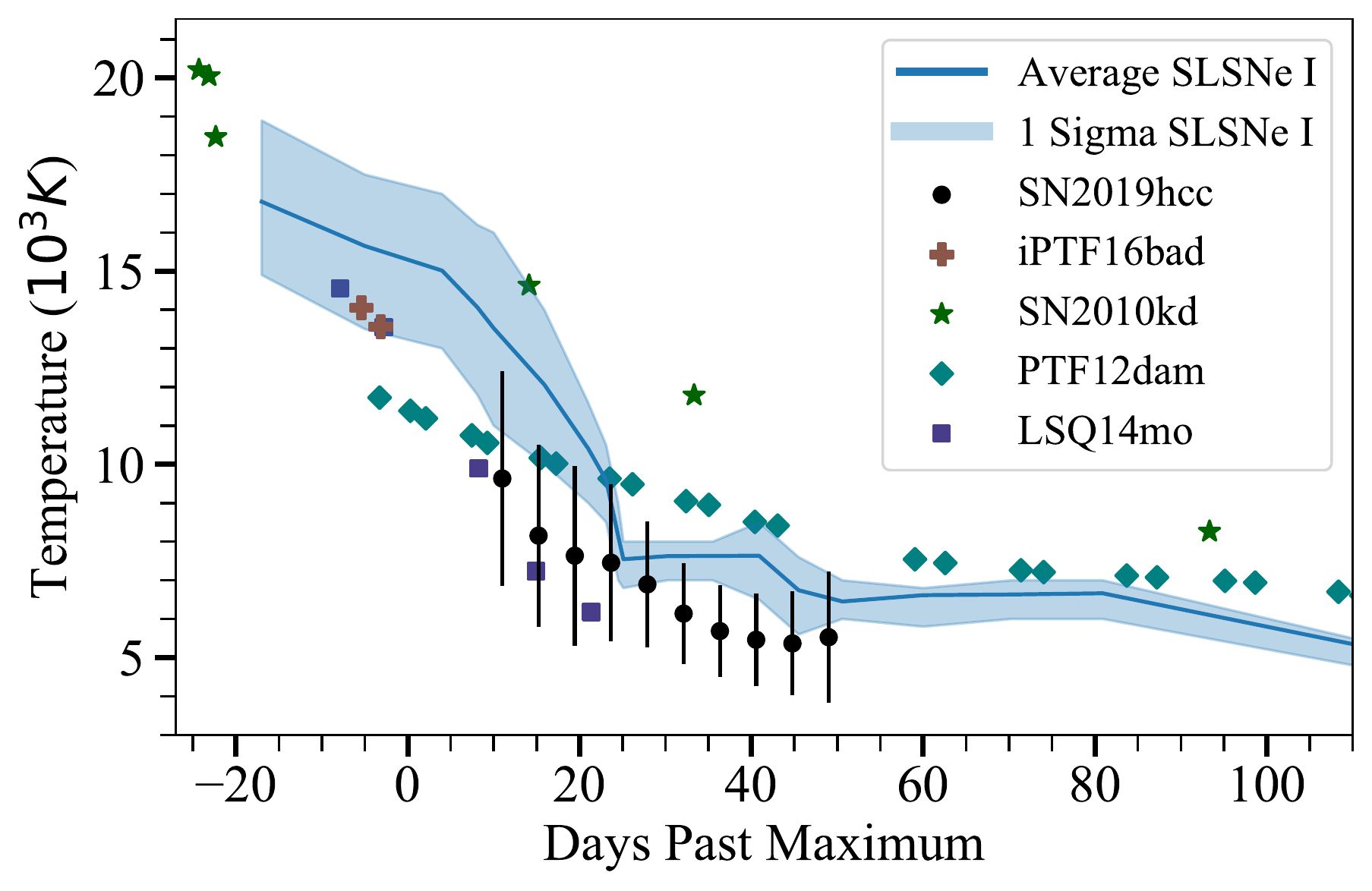}
\vspace{-1mm}
\caption{Temperature comparison of SN~2019hcc with a small sample of SLSNe~I. The closest SLSN~I in temperature to SN~2019hcc at the epoch of +7~days is LSQ14mo. The SLSN temperatures are taken from the literature (see text). The average temperature for SLSNe~I is taken from \protect\cite{Inserra_2017} and reference therein.}
\label{fig: SLSN_temp}
\end{figure}

As the `w' shape profile of SN~2019hcc first spectrum is similar to that observed in SLSNe~I, in Figure \ref{fig: SLSN_temp} we also compare the temperature evolution of SN~2019hcc with a sample of SLSNe~I: iPTF16bad \citep{halpha_2},  SN~2010kd \citep{Kumar_2020}, PTF12dam \citep{Nicholl_2013}, and LSQ14mo \citep{Chen_2017c,Leloudas_2015b}. We selected this small subset of SLSNe~I mainly due to the spectral similarity, see Section~\ref{OII lines} for further information. We also compare to an average temperature evolution for SLSNe~I \citep[][and reference therein]{Inserra_2017}, similarly to what was previously done with SNe~II.
LSQ14mo is the only SLSNe~I with a similar temperature evolution to SN~2019hcc. % at the epoch where the {O~\sc{ii}} lines are observed \citep{Chen_2017c}, and this has the most similarity to SN~2019hcc.

%%%%%%%%%%%%%%%%%%%%%%%%%%%%%%%%%%%%%%%%%%%%%%%%%%%%%%%%%%%%%%%%%%%%%%%%%%%%%%%%
%-----------------------------------------------------------------

\section{Light Curve Analysis}
\label{Light_Curve_Analysis}

\subsection{Bolometric light curve}

We created a pseudo-bolometric light curve from an SED fit to the available photometry, which was interpolated according to the chosen reference band. We used the SDSS $r-$band and the ATLAS $o-$ band as reference, as these bands should approximately cover a similar region of the electromagnetic spectrum, to cover as many epochs as possible. Each band was integrated using the trapezium rule. % - the bands used were $BVgriz$ with the $gri-$bands combining LCO and LT data. 
The redshift, distance, and reddening used were reported in Section \ref{Observations}.

The light curve evolution of SNe~II was considered quantitatively by \cite{Anderson_2014b} and \cite{Valenti_2016}. The decline of the initial steeper slope of a light curve and the second shallower slope can be described as S1 and S2 respectively - in SNe~IIL these are very similar or the same \citep{Anderson_2014b}. S1 and S2 were originally described for $V-$band, however \cite{Valenti_2016} also performs this analysis for pseudo-bolometric light curves and the key parameters are very similar - and in fact the transition between the early fast slope S1 and the shallow late slope S2 is more evident in pseudo-bolometric curves \citep{Valenti_2016}. S2 is followed by the plateau-tail phase \citep{Utrobin_2007}, also known as the post-recombination plateau \citep{book}, which drops into the $^{56}$Co tail. The formalism reported in \cite{Valenti_2016} can be described by the following equation:

\begin{equation}
f(t) = \frac{-A_{\rm 0}}{1+e^{(t-t_{\rm pt})/w_{\rm 0}}} + (t \times p_{\rm 0}) + m_{\rm 0}
\label{eqn 8}
\end{equation}

Here the variables $A_{\rm 0}, w_{\rm 0}, m_{\rm 0}$ are free parameters describing the shape of the drop, $p_{\rm 0}$ describes the decline of the tail, and $t_{\rm pt}$ describes the length of the plateau, measured from the explosion to the midpoint between the end of the plateau phase and start of the radioactive tail.

\begin{figure}
\centering
\includegraphics[width=1\linewidth]{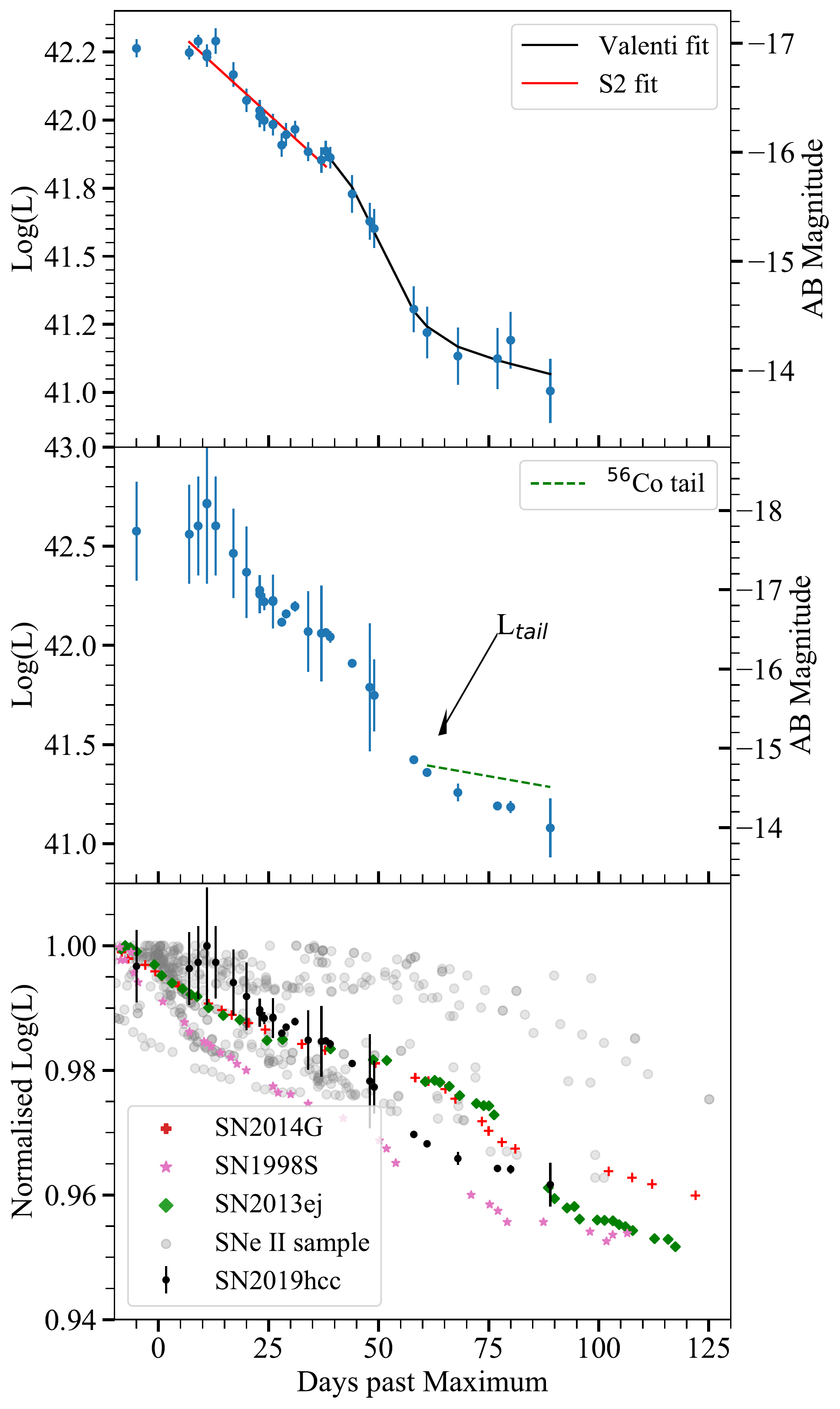}
\vspace{-1mm}
\caption{Top panel: the pseudo-bolometric light curve of SN~2019hcc, with $r-$band as the reference. Middle panel: the bolometric light curve of SN~2019hcc. The tail magnitude and comparison to $^{56}$Co decay rate is marked. Bottom panel: the bolometric light curve of SN~2019hcc, compared to those of SNe~II SN~2014G \protect\citep{Terreran_2016}, SN~2013ej \protect\citep{Huang_2015}, and SN~1998S \protect\citep{Fassia_2000,Fassia_2001}. The bolometric light curves of the sample of SNe~II from \protect\cite{Faran_2014} is shown in grey - two distinct branches can be seen which could be described with the SN~IIL and SN~IIP subcategories. The light curves have been normalised with respect to the maximum.}
\label{fig: Bolometric}
\end{figure}

The top panel on Figure \ref{fig: Bolometric} shows the pseudo-bolometric light curve, % - a `plateau'-like slope was identified, 
however there is no distinguishable change in the slope leading to a clear distinction of S1 and S2, and after approximately 60 days past maximum the light curve transits into a `plateau-tail phase' and then drops into a radioactive tail. As there are not multiple slopes in the initial decline, S1 and S2 will hereafter be collectively referred to as S2 for SN~2019hcc, leading to a Type IIL sub-classification for the supernova. The S2 decline was found to be $1.51\pm0.09$ mag per 50 days. The best-fit $t_{\rm pt}$ was $66.0\pm1.1$ days, and $p_0$ was measured via a linear fit and found to be 
%$1.01\pm0.98$ mag per 100 days. This is a large error so the decline was also measured with a linear fit to match the Valenti fit, giving a value of
$1.38\pm0.49$ mag per 100 days. \cite{Valenti_2016} found a mean length of the plateau in SNe~II of $t_{\rm pt} = 100$, which is up to the midway in the plateau-tail phase. Considering this average, SN~2019hcc has a relatively short plateau duration, which could suggest a lower ejecta mass, but could also be due to a smaller progenitor radius or a higher explosion energy \citep{Popov_1993}. This fitting was performed for the pseudo-bolometric light curve rather than $V-$band due to the sparsity of photometric data in this band, particularly for the tail of the light curve.

The middle panel of Figure \ref{fig: Bolometric} shows the full bolometric light curve - this was found by fitting a blackbody to the photometry and integrating between 200~$\Angstrom$ and 25000~$\Angstrom$. The bolometric light curve required interpolation and extrapolation of additional points for epochs where some bands were not observed. This was done by taking a constant colour from the nearest points in the other bands - however this is an assumption which increases the uncertainty in the resultant curve. The tail luminosity $L_{\rm tail}$ is marked, and a $^{56}$Co tail has been plotted using Equation \ref{eqn 9}, as from \cite{Jerkstrand_2012}, which gives the bolometric luminosity for the theoretical case of a fully-trapped $^{56}$Co decay.

If full trapping of gamma-ray photons from the decay of $^{56}$Co is assumed, the expected decline rate is 0.98~mag per 100 days in $V-$band \citep{Woosley_1989,Anderson_2014b}. The tail of SN~2019hcc clearly declines faster than the $^{56}$Co tail as shown in the middle panel of Figure \ref{fig: Bolometric}. If this is indeed the radioactive tail, it seems that SN~2019hcc displays incomplete trapping. This is not entirely unexpected as \cite{Gutierrez_2017b} showed that most fast-declining SNe show a tail decline faster than expected from $^{56}$Co decay. \cite{Terreran_2016} found incomplete trapping for SN~2014G, one of the SNe in our comparison sample. They suggested a few possibilities for incomplete trapping such as a low ejecta mass, high kinetic energy or peculiar density profiles. However, dust formation could also result in a fast-declining tail, and additional effects such as a different radioactivities could affect the decline \citep{book}, as well as CSM-ejecta interaction, which can contribute to the luminosity at late times \citep[e.g.][]{Andrews_2019}.

The lower panel of Figure \ref{fig: Bolometric} shows a comparison of the bolometric light curve of SN~2019hcc with Type~II SNe 2013ej and 2014G, and with the Type~IIn SN~1998S. These were chosen for comparison as they present a similar photometric evolution to SN~2019hcc (see Section \ref{Photometry}). The bolometric light curves from the sample of SNe~II from \cite{Faran_2014} are also included, and two distinct branches can be seen which would correspond to the historic SN~IIL and SN~IIP sub-classifications. However, note the small sample size of this study compared with other sample analyses. All light curves have been normalised by the peak luminosity for comparison. This panel supports that the sample of SNe~II discussed would all be considered SNe~IIL, or fast decliners. 

\begin{table}
\begin{tabular}{ c c c c c }
\hline 
SN &V$_{\rm 50}$ & Rise (days) & Peak (Absolute Mag) \\ 
\hline 
SN~2019hcc & 1.52$\pm$0.03 & 15.3$\pm$7.4 & $-17.7$ \\
SN~2014G & 1.58$\pm$0.06 & 14.4$\pm$0.4 & $-18.1$ \\
SN~2013ej & 1.24$\pm$0.02 & 16.9$\pm$1 & $-17.64$\\
SN~1998S & 1.87$\pm$ 0.07 & $\sim$18 & $\sim -18.1$ \\
\multirow{2}{*}{SNe~II} & 1.43$\pm$0.21 (IIL) & \multirow{2}{*}{$16.0\pm$3.6} & \multirow{2}{*}{$-16.96\pm$1.03}\\ 
& 0.31$\pm$0.11 (IIP) \\

\hline 
\end{tabular}
\caption{Here V$_{\rm 50}$ is the $V-$band mag decline in the first 50 days (roughly equivalent to S2), measured directly from the light curves with a linear fit. Rise times and peak absolute magnitude are in $R-$band for SN~2014G \protect\citep{Terreran_2016} and SN~2013ej \protect\citep{Richmond_2014,Huang_2015}, or ATLAS $o-$band for SN~2019hc. Rise time and peak values for SN~1998S are also in the $R-$band, however they are estimated from the light curve rather than taken from literature. Also shown are the average values for a sample of 10 SN~IIL and 18 SN~IIP from \protect\cite{Faran_2014}. Though these populations have been previously discussed as continuous, the distinction is still useful to give context to the measured values. \protect\cite{Anderson_2014b} found a mean S2 of 0.64 for a sample of 116 SNe~II, roughly the average of the IIL and IIP sub-classes in the above. The rise time for SNe~II is taken from \protect\cite{Pessi_2019}. The average absolute peak magnitude in $R-$band from SNe~II comes from \protect\cite{Galbany_2016}.}
\label{tab: sn_prop}
\end{table}

%SN~2014G $\sim$4.8-7 
%SN~2013ej $\sim$10.6
%Rise time and absolute mag are in $R-$band, or ATLAS $o-$band for SN~2019hcc. SN~2014G \citep{Terreran_2016}, SN~2013ej \citep{Richmond_2014,Huang_2015}

A SN~IIL has been defined as where the $V-$band light curve declines by more than 0.5~mag from peak brightness during the first 50 days after explosion \citep[e.g.][]{Faran_2014}. The initial decline of SN~2019hcc was also measured in $V-$band and is displayed, along with other properties, in Table \ref{tab: sn_prop} together with the comparison SNe and the average values for SN~IIP and SN~IIL. Looking at Figure \ref{fig: Bolometric}, the S2 slope of SN~2019hcc appears steeper, and the plateau shorter, than the comparison SNe~II SN~2014G and SN~2013ej. However, SN~1998S has a faster intial decline, and appears to transition to the tail at a comparable epoch. SN~2013ej has the most distinct S1 and S2. The radioactive tail of SN~2019hcc shows a similar decline rate to all comparison SNe which also seem to display incomplete trapping, or at the very least a radioactive tail decay faster than $^{56}$Co decay. The SN~2019hcc light curve evolution drops out of the photospheric phase sooner than SN~2013ej and SN~2014G - implying a lower ejecta mass. It could therefore be suggested that the ejecta mass of SN~2019hcc is lower than the that of these other SNe, however other factors such as explosion energy could also play a role \citep{Popov_1993}. 

%\cite{Popov_1993} gives the following equation to calculate ejecta mass:

%\begin{equation}
%\rm log(M) = 4log10(t_{\rm ph}) + 0.4V + 3log10(v_{\rm ph})-2.089
%\label{popov}
%\end{equation}

%Where $M$ is the ejecta mass in M$_{\odot}$, $t_{\rm pt}$ is the length of the plateau in days, $v_{\rm pt}$ is the photospheric velocity in 10$^3$~km/s of the plateau, and $V$ is the absolute V-band magnitude of the plateau. As these values are not constant, they were measured for the start, middle and end of the plateau and averaged to give 2.7$\pm$0.7~M$_{\odot}$. {\bf need to redo as was considering wrong tp}.
%{\bf WHAT ABOUT 1998S?}

%%%%%%%%%%%%%%%%%%%%%%%%%%%%%%%%%%%%%%%%%%%%%%%%%%%%%%%%%%%%%%%%%%%%%%%%%%%%%%%%
\subsection{$^{56}$Ni Production}
\label{Ni mass}

\cite{Jerkstrand_2012} presented a method to retrieve the $^{56}$Ni mass produced by comparing the estimated bolometric luminosity in the early tail-phase with the theoretical value of fully trapped $^{56}$Co deposition, which is given by:

\begin{equation}
L(t) = 9.92\times10^{41} \times \frac{M_{^{56}\rm Ni}}{0.07M_{\odot}} \times (e^{ -t/111.4 d} - e^{ -t/8.8d})
\label{eqn 9}
\end{equation}

Where $t$ is the time since explosion, $L(t)$ is the luminosity in ergs$^{-1}$ at that time, 8.8~days is the e-folding time of $^{56}$Ni and 111.14~days is the e-folding time of $^{56}$Co decay. It is also assumed that the deposited energy is instantaneously re-emitted and that no other energy source has any influence. To calculate the mass of $^{56}$Ni, the tail luminosity and the time at which the tail begins should be used in Equation~\ref{eqn 9}.

A visible transition can be seen in Figure \ref{fig: Bolometric} into the tail of SN~2019hcc at 61 days past maximum, therefore we selected the tail luminosity as the magnitude at the point of transition. With this tail magnitude, according to the above approach, the mass of $^{56}$Ni is $0.035\pm0.008$~M$_\odot$. The uncertainty was calculated as 0.1 dex, as a measure of the distance to the adjacent points, as the exact location of the tail start is uncertain. This is only a lower limit due to likely incomplete trapping. \cite{Anderson_2014b} performed this analysis on a large set of SNe~II, and found a range of $^{56}$Ni masses from 0.007 to 0.079~M$_\odot$, with a mean value of 0.033~M$_\odot$ ($\sigma$= 0.024). A survey of literature values led to a mean mass $^{56}$Ni~=~0.044~M$_\odot$  for a sample of 115 SNe II \citep{Anderson_2019}. Therefore we conclude that the value retrieved for SN~2019hcc is within the expected range for a SN~II.

%%%%%%%%%%%%%%%%%%%%%%%%%%%%%%%%%%%%%%%%%%%%%%%%%%%%%%%%%%%%%%%%%%%%%%%%%%%%%%%%
%-----------------------------------------------------------------

\section{Spectroscopy}
\label{Spectroscopy}

\begin{figure}
\centering
\includegraphics[width=1\linewidth]{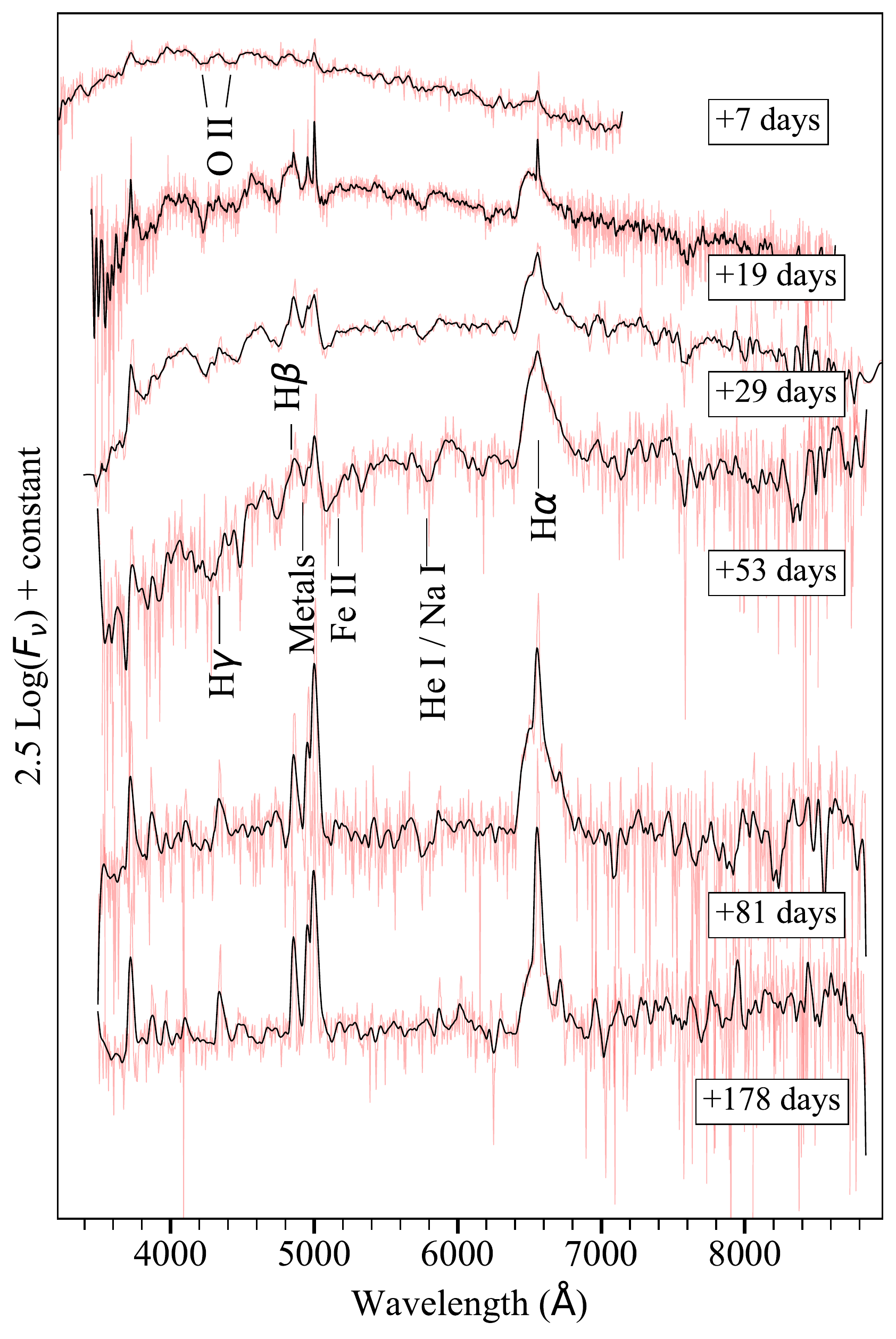}
\vspace{-1mm}
\caption{The spectra for SN~2019hcc and their phase with respect to maximum light (MJD~58636). The wavelength is in the rest frame. The spectra have been corrected according to the photometry (excluding the last epoch which had no photometry available), de-reddened and redshift-corrected. They have also been smoothed using a moving average  - this recalculates each point as the average of those on either side, in this case for five iterations  - (black) with the original overlaid (red). The flux has been converted to log($F_{\nu}$) to emphasize absorption features. The most prominent elements have been labelled - here `Metals' refers to a combination of {Ba~\sc{ii}}, {Sc~\sc{ii}} and {Fe~\sc{ii}}.}
\label{fig: spectra}
\end{figure}

Figure \ref{fig: spectra} shows the spectral evolution of SN~2019hcc, labelled with the phase with respect to maximum light (MJD~58636). The spectra have been flux-calibrated according to the broadband photometry. The last epoch was not calibrated according to the photometry as none was available. At +81~days, the SED no longer follows a blackbody assumption as the ejecta is now optically thin and the photospheric phase is over, however the blackbody fit to the the photosphere is a valid approximation for the earlier spectra. The light curve analysis from Section~\ref{Light_Curve_Analysis} suggests the end of the plateau/photospheric phase, $t_{\rm{pt}}$, at approximately +66 days from explosion. %The spectra support this as by +81~days SN~2019hcc no longer follows a blackbody continuum, suggesting the end of the photospheric phase. %- and \cite[e.g][]{Anderson_2014b} measures the duration of the photospheric phase at the end of the plateau.
Emission lines from the host galaxy can be seen, particularly from +53~days. The resolution of the spectra can be found in Appendix \ref{Appendix: Data}, Table \ref{Resolution_table}.

%The spectra have been mangled according to the blackbody fits to the closest photometry available to the epoch of the spectra.

The spectra were also corrected for redshift and de-reddened according to the Cardelli Extinction law using Av = 0.19 mag and Rv~=~3.1 \citep{Cardelli_1989}. They have been offset for clarity on an arbitrary y-axis. The flux has been converted to log($F_{\nu}$) where $F(\nu) = F(\lambda)\lambda^2/3e18$ to highlight the absorption features. 

As can be seen, the first spectrum at +7~days after peak displays a `w'-shaped profile at the rest-wavelengths typical of {O~\sc{ii}} lines with absorption minima at approximately 4420~$\Angstrom$ and 4220~$\Angstrom$, which originally motivated the classification as a SLSN I. However, these signatures disappear in subsequent spectra with the H$\alpha$ emission becoming the dominant spectral feature. Aside from the w-shape, the first spectrum is relatively featureless. A well developed H$\alpha$ profile can be seen from +19~days, as well as H$\beta$ and H$\gamma$, though less developed Balmer lines can also be seen at +7~days. {Fe~\sc{ii}} and {He~\sc{i}} lines can also be seen from the +7~days spectrum and become well-developed by +19~days. The typical core-collapse SN forbidden lines of [{O~\sc{i}}] at $\lambda\lambda$6300, 6363  and [{Ca~\sc{ii}}] at $\lambda\lambda$7291, 7323 are not seen despite SN~2019hcc appearing to reach the nebular phase, which roughly starts at 100-200 days \citep{Chevalier_1989}. There could be a few possibilities for their absence. The first is that the nebular phase has not been reached. Alternatively, as the strength of [{O~\sc{i}}] increases with the ZAMS mass \citep[e.g.][]{Dessart_2020}, it would imply a ZAMS mass of the SN~2019hcc progenitor sufficiently low that the [{O~\sc{i}}] are not visible. Another possibility is that SN~2019hcc is too faint with respect to the host and the lines have not yet developed.

In SN~2014G, after $\sim$80 days the emission feature of [{Ca~\sc{ii}}] at $\lambda\lambda$7291, 7323 starts to become visible, approximately coincident with the sudden drop in the light curve \citep{Terreran_2016}. SN~2013ej also shows [{Ca~\sc{ii}}] and [{O~\sc{i}}] forbidden lines from 109~days, when the SN entered the nebular phase \citep{Bose_2015}, suggesting SN~2019hcc is unusual in this respect. However, \cite{book} noted the spectra of some SN~IIL (e.g. SN~1986E, SN~1990K) do not contain the standard emission lines of core-collapse supernovae, and the forbidden lines arising in the ejecta may be suppressed by high densities or obscured by the circumstellar medium (CSM) that produces the extended Hydrogen̨ emission.

The flux of H$\alpha$ in the +178~days spectrum (excluding the narrow host contribution component) is $\sim$5 times that of H$\beta$. For case B recombination in the temperature regime 2500 $\leq$ T(K) $\leq$ 10000 and electron density 10$^2$ $\leq$ $n_{e}$ $\leq$ 10$^6$, the H$\alpha$ line should be 3 times stronger than H$\beta$ \citep{Ferland_2006}. However, the case B recombination is not observed in SNe~II before a couple of years. \cite{Kozma_1998} suggested at 200 days past explosion in SN~1987A this ratio should have been around 5, based on the total calculated line flux and using a full Hydrogen atom with all nl-states up to $n=20$ included. The ratio of SN~2019hcc appears similar to SN~1987A and other SNe~II at the onset of the nebular phase. Despite the H$\alpha$/H$\beta$ ratio being higher than the case B recombination, it is still sufficiently low that we can conclude that any additional flux to H$\alpha$ should be insignificant. Excess flux in H$\alpha$ could be a clue that H$\alpha$ is also collisionally excited, suggesting interaction \citep{Branch_1981}. As the H$\alpha$ profile evolves it appears to become asymmetrical, suggesting a multi-component fit in the late spectra. The simplest explanation for this is that a mostly spherical ejecta is interacting with a highly asymmetric, Hydrogen-rich CSM \citep{Benetti_2016}. This is in contrast with the quick decay of the tail, suggesting that such asymmetry might be intrinsic of the ejecta or the result of other lines that are not resolved, for example [{N~\sc{ii}}] $\lambda$6584. An asymmetric line profile can also be interpreted as evidence for dust formation in the ejecta \citep[e.g.][]{Smith_2008}.

%%%%%%%%%%%%%%%%%%%%%%%%%%%%%%%%%%%%%%%%%%%%%%%%%%%%%%%%%%%%%%%%%%%%%%%%%%%%%%%%

\subsection{Spectral Comparison }
\label{Spectra_Comp}
\begin{figure}
\centering
\includegraphics[width=1\linewidth]{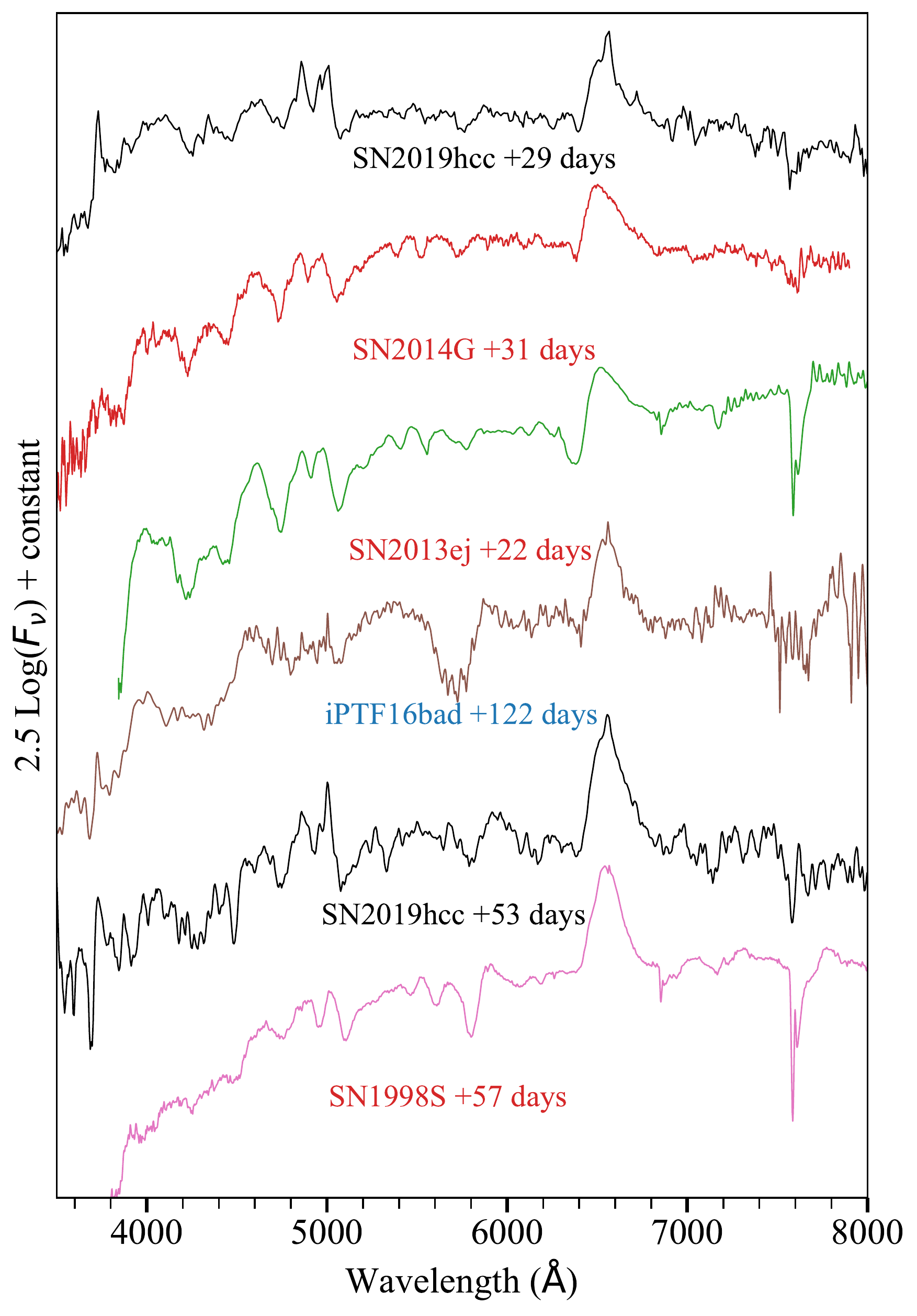}
\vspace{-1mm}
\caption{SN~2019hcc at +29~days post peak is compared to moderately luminous Type II SN~2014G and SN~2013ej, and SLSN~I iPTF16bad which displays H$\alpha$ at late time (at +100~days post peak) in its spectra. SN~2019hcc at +53~days past peak is also compared to SN~1998S. The wavelength is in the rest frame. Text in red refers to Type II, while in blue to the only SLSN~I.}
\label{fig: spectra_comp}
\end{figure}

Comparison of SN~2019hcc with the moderately luminous SNe~II \citep{moderates} reported in Section~\ref{Light_Curve_Analysis} and Figure~\ref{fig: blackbody}, together with SLSN~I iPTF16bad, is shown in Figure \ref{fig: spectra_comp} \footnote{These spectra were taken from  the Weizmann Interactive Supernova Data Repository (WISeREP) \citep{wiserep}}.

iPTF16bad at late times displays H$\alpha$ emission due to the collision with a H-shell ejected approximately 30 years prior, thought to be due to pair instability pulsations \citep{halpha_1,halpha_2}, and merits comparison as it is a SLSN~I displaying a w-shaped profile at early times and H$\alpha$ at late times. SN~2019hcc has a good match with some features, e.g. Balmer lines, however there are some discrepancies in the comparison, such as the lack of a P-Cygni profile for H$\alpha$ in iPTF16bad. The {Fe~\sc{ii}} lines at approximately 5000~$\Angstrom$ are also not observable in the spectrum of the SLSN~I. If the H$\alpha$ in SN~2019hcc was a consequence of interaction similar to iPTF16bad, we would expect other signs of interaction. These could be undulations or a second peak in the light curve \citep[e.g.][]{Nicholl_2016,Inserra_2017}, but the SN~2019hcc light curve appears to be that of a typical SN~IIL (see Section~\ref{Light_Curve_Analysis}). Additionally, the relatively earlier appearance of the H$\alpha$ emission in SN~2019hcc would require a much closer H-shell than for iPTF16bad.

%At early times, the temperature in the ejecta could mean almost all the hydrogen present is ionised, and therefore this hydrogen is not visible in the spectra. Therefore it is not clear whether the hydrogen is also present at earlier times and simply not visible.

Figure~\ref{fig: spectra_comp} also displays a comparison to moderately luminous SNe~II SN~2013ej, SN~2014G, and SN~1998S. SN~1998S did not have a spectrum available at the +29~days epoch, so is shown at the nearest later epoch, with SN~2019hcc at +53~days for comparison. The spectra do not significantly evolve in this time frame. There are strong similarities between spectral features at the epoch of comparison, %In fact the resemblance is much more significant, 
with good matches of H$\alpha$ and {Fe~\sc{ii}} features. The comparison would strengthen that SN~2019hcc is a Type II. 

\begin{figure}
\centering
\includegraphics[width=1\linewidth]{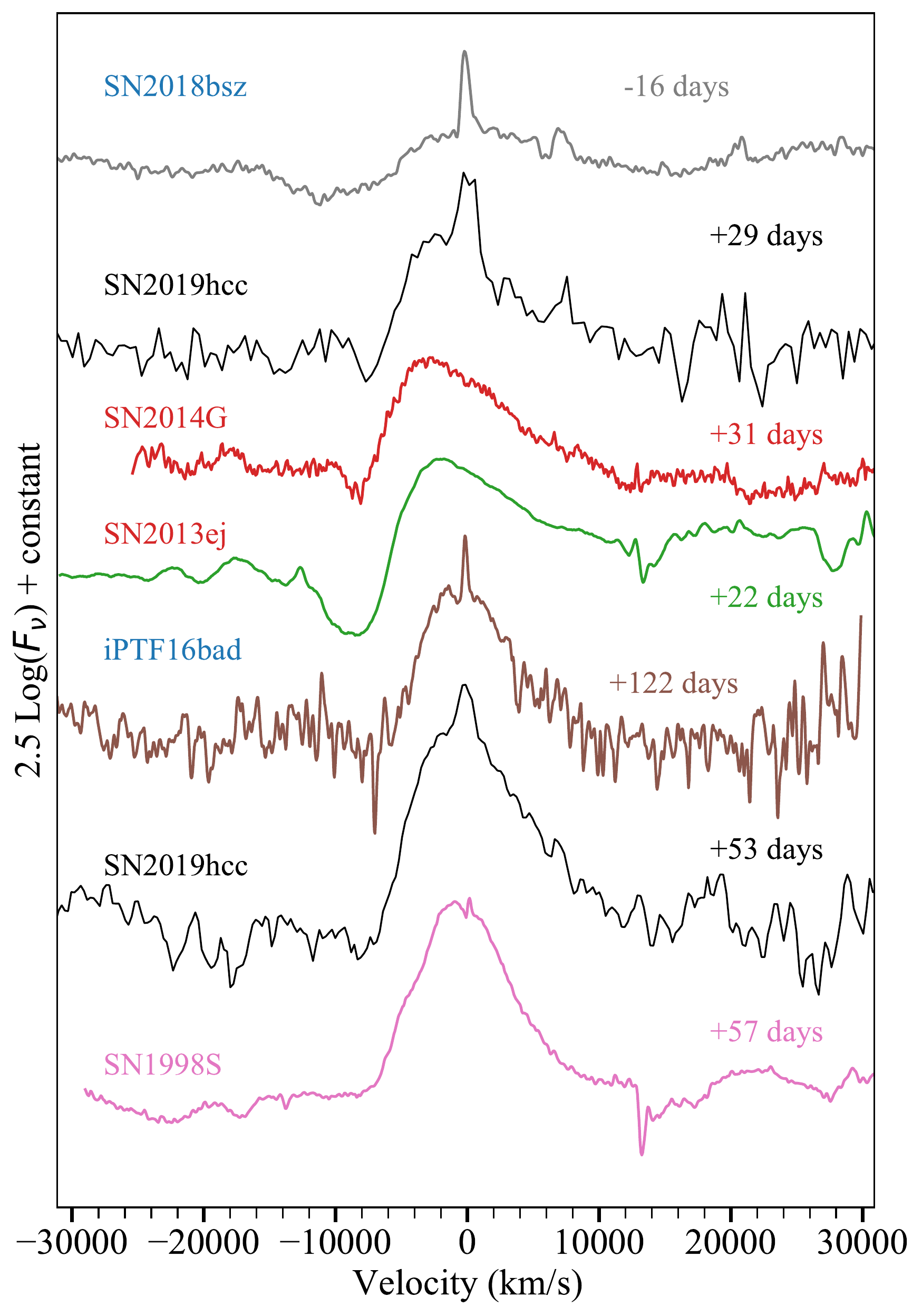}
\vspace{-1mm}
\caption{A comparison of the H$\alpha$ profiles for a variety of moderately luminous SNe~II and SLSNe~I. The wavelength is in the rest frame. Text in red refers to Type II, while in blue to SLSNe~I. The velocity is with respect to H$\alpha$.}
\label{fig: halpha_comp}
\end{figure}

Figure \ref{fig: halpha_comp} shows a closer look at the H$\alpha$ profiles for the previous spectra, and additionally SN~2018bsz, a SLSN~I. In SLSNe~I, carbon lines produced in the H$\alpha$ region could be mistaken for Hydrogen, such as in the case of SN~2018bsz, which displays {C~\sc{ii}} $\lambda$6580 line in the H$\alpha$ region \citep{2018bsz}. SN~2018bsz does also show Hydrogen but it is not observed at the phase being considered here. However, if {C~\sc{ii}} is present in a spectrum, we should observe it at $\lambda$7234 and $\lambda$5890 \citep{2018bsz}, lines which are not seen in SN~2019hcc, while H$\beta$ can be seen at $\lambda$4861. This strengthens the idea that is indeed H$\alpha$ observed in SN~2019hcc as opposed to {C~\sc{ii}}.

SN~2019hcc spectra show an emission redward of H$\alpha$ at approximately 6720~$\Angstrom$ visible at +29~days.
Figure \ref{fig: halpha_comp} shows that SN~2018bsz also contains the redward emission at approximately 6720~$\Angstrom$. \cite{Singh_2019} identifies this as [{S~\sc{ii}}] lines at 6717~$\Angstrom$ and 6731~$\Angstrom$ from the parent {H~\sc{ii}} region.

%%%%%%%%%%%%%%%%%%%%%%%%%%%%%%%%%%%%%%%%%%%%%%%%%

%\subsection{High Velocity Components}
\subsection{Investigating Signs of Interaction in the Photospheric Spectra}

A multi-component H$\alpha$ profile which does not completely hide the absorption component hints to a degree of interaction. Here the narrow component would belong to the unshocked wind, whilst the medium component to the shocked wind/ejecta. Another sign of interaction between the ejecta and the CSM could be a high velocity (HV) component in the Balmer lines \citep[e.g.][]{moderates, Gutierrez_2017a}. The normal velocity originates from the receding photosphere, whilst the high velocity is generated further out where the CSM interaction may excite the Hydrogen to cause a second, high-velocity absorption feature \citep[e.g.][]{SNIIbook}. The size and shape of this feature could be related to the progenitor wind density \citep{Chugai_2007}. A small absorption bluer than the H$\alpha$ P-Cygni has been observed in several SNe~II but its nature is not always linked to H$\alpha$ \citep{Gutierrez_2017a}. Such a feature, named `Cachito', has previously been attributed to HV features of Hydrogen, or {Si~\sc{II}} $\lambda$6533. These features were identified in \cite{moderates} for some moderately luminous SNe~II. \cite{Gutierrez_2017a} also found the `Cachito' feature is consistent with {Si~\sc{ii}} at early phases, and with Hydrogen at later phases.

\begin{figure}
\centering
\includegraphics[width=1\linewidth]{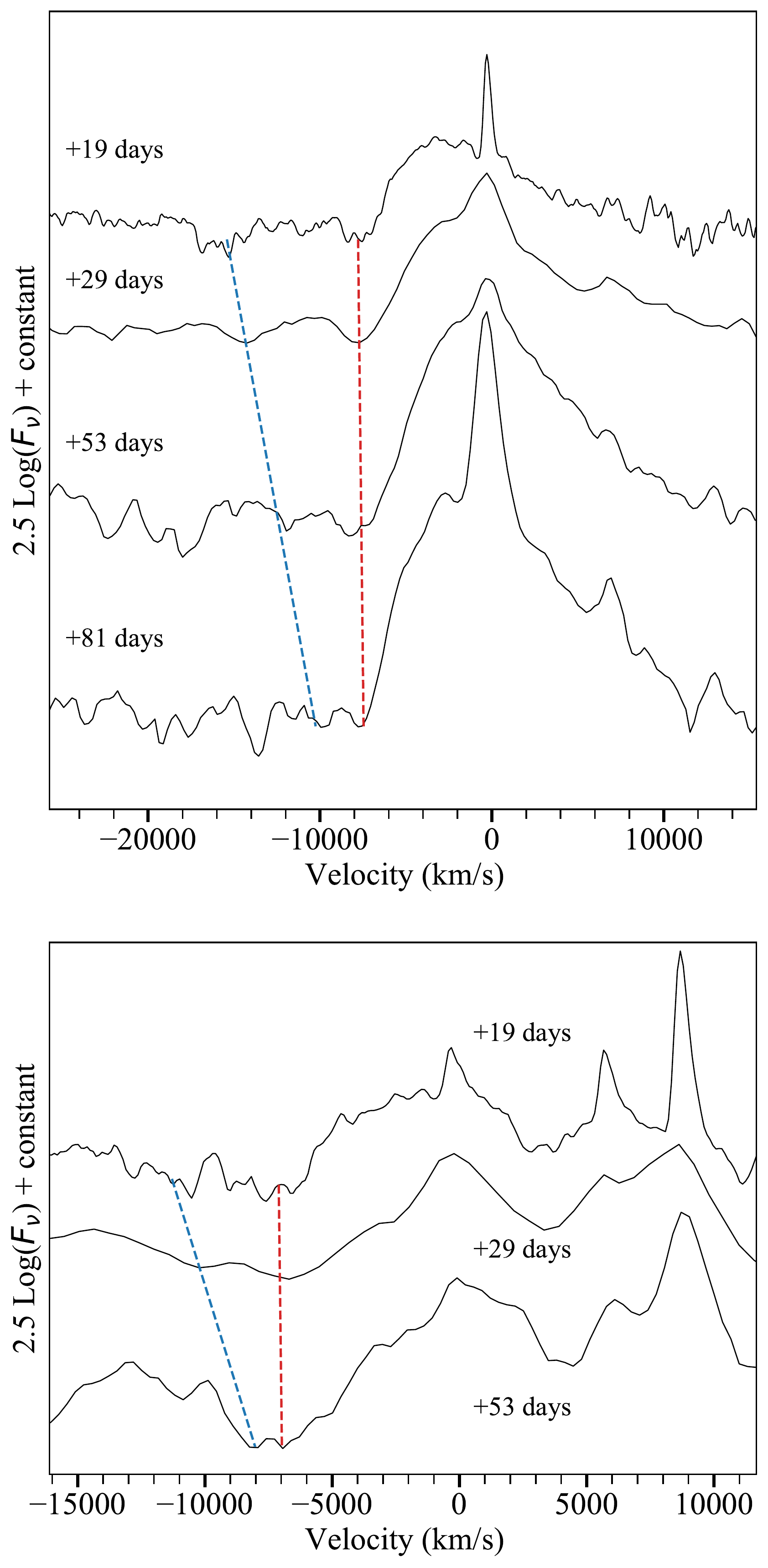}
\vspace{-1mm}
\caption{Top panel: the H$\alpha$ profile evolution of SN~2019hcc, the spectra have been smoothed using a moving average. The velocity is with respect to H$\alpha$. All spectra show a small feature blue-ward of H$\alpha$ after smoothing, which could be a HV component indicating early CSM-ejecta interaction. The red dashed line tracks the H$\alpha$ absorption, and the blue dashed line the possible HV component. Bottom panel: the same as the above panel but with respect to H$\beta$. The velocity is with respect to H$\beta$.}
\label{fig: HV}
\end{figure}

\subsubsection{High Velocity features}\label{sec:hv}
In the top panel of Figure \ref{fig: HV}, an absorption blue-ward of H$\alpha$ can be seen in SN~2019hcc at +19~days and +29~days at around 6250~$\Angstrom$, %confirming this is a real feature. There is a clear absorption blueward of H$\alpha$ up to +29~days, 
however after this epoch it becomes less clear. An absorption feature can also be seen in SN~2013ej, and arguably SN~2014G, as seen in Figure \ref{fig: halpha_comp}.
The presence of a potential HV H$\beta$ additional to the H$\alpha$ at a similar velocity would strengthen the latter's status as a HV feature of Hydrogen \citep[e.g.][]{Chugai_2007,Gutierrez_2017a,Singh_2019}. The lower panel on Figure \ref{fig: HV} shows the H$\beta$ profile for SN~2019hcc at the epochs where it is visible, and an absorption blueward of the P-Cygni could be identified. \cite{Gutierrez_2017a} found that  63\% of their sample of SNe~II with HV H$\alpha$ in the plateau phase showed a HV H$\beta$ at the same velocity. \cite{Gutierrez_2017a} also reported that if the absorption is produced by {Si~\sc{ii}} its velocity should be similar to those presented by other metal lines, such as {Fe~\sc{ii}} $\lambda5169$, a good estimator for the photospheric velocity \citep{Hamuy_2001}.

%Velocity evolution can give an indication of the explosion geometry. During the plateau phase {Fe~\sc{ii}} $\lambda5169$ is a good estimator for the photospheric velocity, and at early times {He~\sc{i}} or H$\beta$ are good tracers \citep{Hamuy_2001}. The relatively higher velocities of H$\alpha$ and H$\beta$ at all phases compared to {Fe~\sc{ii}} $\lambda5169$ would suggest formation at larger radii \citep{Bose_2015}.

\begin{figure}
\centering
\includegraphics[width=1\linewidth]{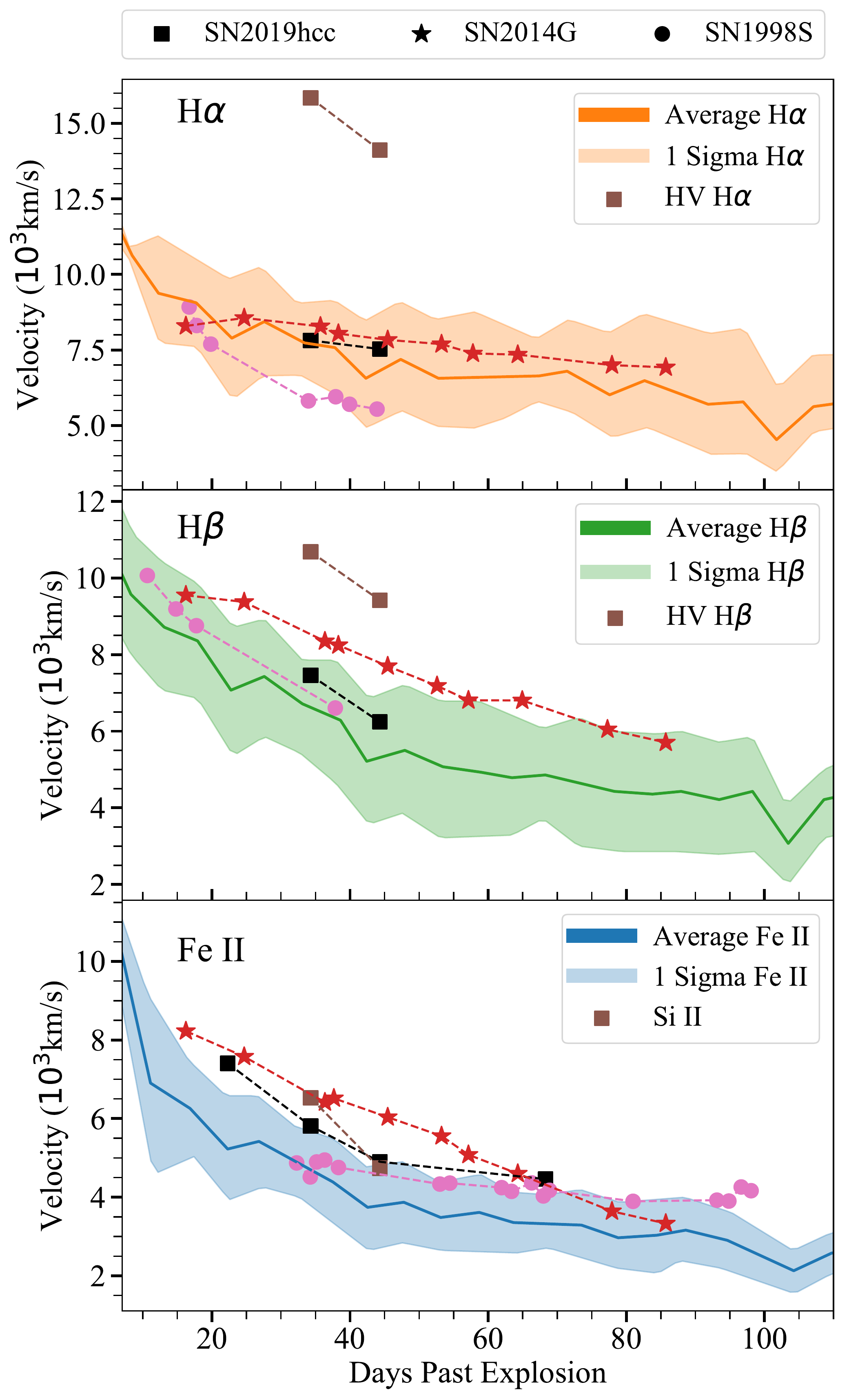}
\vspace{-1mm}
\caption{The velocity comparisons of different lines for SN~2019hcc, over three different epochs, alongside the velocities of SN~2014G for {Fe~\sc{ii}}, H$\alpha$ and H$\beta$. Velocities of SN~1998S from \protect\cite{Anupama_2001} and \protect\cite{Terreran_2016}, SN~2014G from \protect\cite{Terreran_2016}. The average velocities are found from 122 Type II SNe \protect\citep{Gutierrez_2017a}, the figure reproduced from \protect\cite{Dastidar_2020} and reference therein, shown with a 1-sigma error.}
\label{fig: velocity_comp}
\end{figure}

The velocity of this possible H$\alpha$ HV absorption feature in SN~2019hcc was measured at +19~days and +29~days, with respect to H$\alpha$ and {Si~\sc{ii}} at $\lambda 6355$. The {Fe~\sc{ii}} lines were also measured for comparison. Figure \ref{fig: velocity_comp} displays the measured velocities in SN~2019hcc for various lines at different epochs in its evolution. The velocity was found by fitting a Gaussian to the absorption features and finding the minimum -  after +29~days, this fitting was not successful, therefore there are only two points available. With reference to Figure \ref{fig: velocity_comp} it can be seen that the measured {Si~\sc{ii}} velocity is close to the {Fe~\sc{ii}} velocity at both epochs, suggesting that it is near the photospheric velocity. This would lend support to the feature being more likely associated with {Si~\sc{ii}}. For the HV component, it would be expected the velocity of the HV H$\beta$ to match that of the HV H$\alpha$, and this is not what is found by our velocity analysis. Considering this information, %whilst a HV component of H$\alpha$ and H$\beta$ could possibly be identified in the spectra,
the velocity measurements support that this feature is most likely associated with {Si~\sc{ii}}. 

Velocities were also measured for the lines of H$\alpha$, H$\beta$, and {Fe~\sc{ii}} in SN~2014G and SN~1998S as shown in Figure \ref{fig: velocity_comp}, and the {Fe~\sc{ii}} velocities are similar to those of SN~2019hcc - although due to the scarcity of points for SN~2019hcc a meaningful comparison of the velocity evolution is difficult. Additionally the average velocities of these lines as measured by \cite{Gutierrez_2017a} for a sample of 122 SNe~II are included in the plot, and show the velocities measured for SN~2014G, SN~2019hcc and SN~1998S are roughly as expected for SNe~II.

%A relatively narrow FWHM of 3000–4000~Km/s for the {O~\sc{ii}} absorption could additionally support {O~\sc{ii}} being confined to a thin shell \citep{Quimby_2018}. 

%%%%%%%%%%%%%%%%%%%%%%%%%%%%%%%%%%%%%%%%%%%%%%%%%%%%%%
\subsubsection{Photospheric H$\alpha$ profile}
Another sign of interaction in the spectra would be a multi-component H$\alpha$ profile with additional components to a simple P-Cygni profile. To investigate the possible presence of multi-components, the profile of SN~2019hcc taken from its highest resolution spectrum at +19 days was decomposed by means of Gaussian profiles. In a non-perturbed SN ejecta, the expected components would be both an absorption and an emission from the P-Cygni, as well as emission from the host galaxy. Any additional component could therefore suggest an ongoing ejecta-CSM interaction. 

\begin{figure}
\centering
\includegraphics[width=1\linewidth]{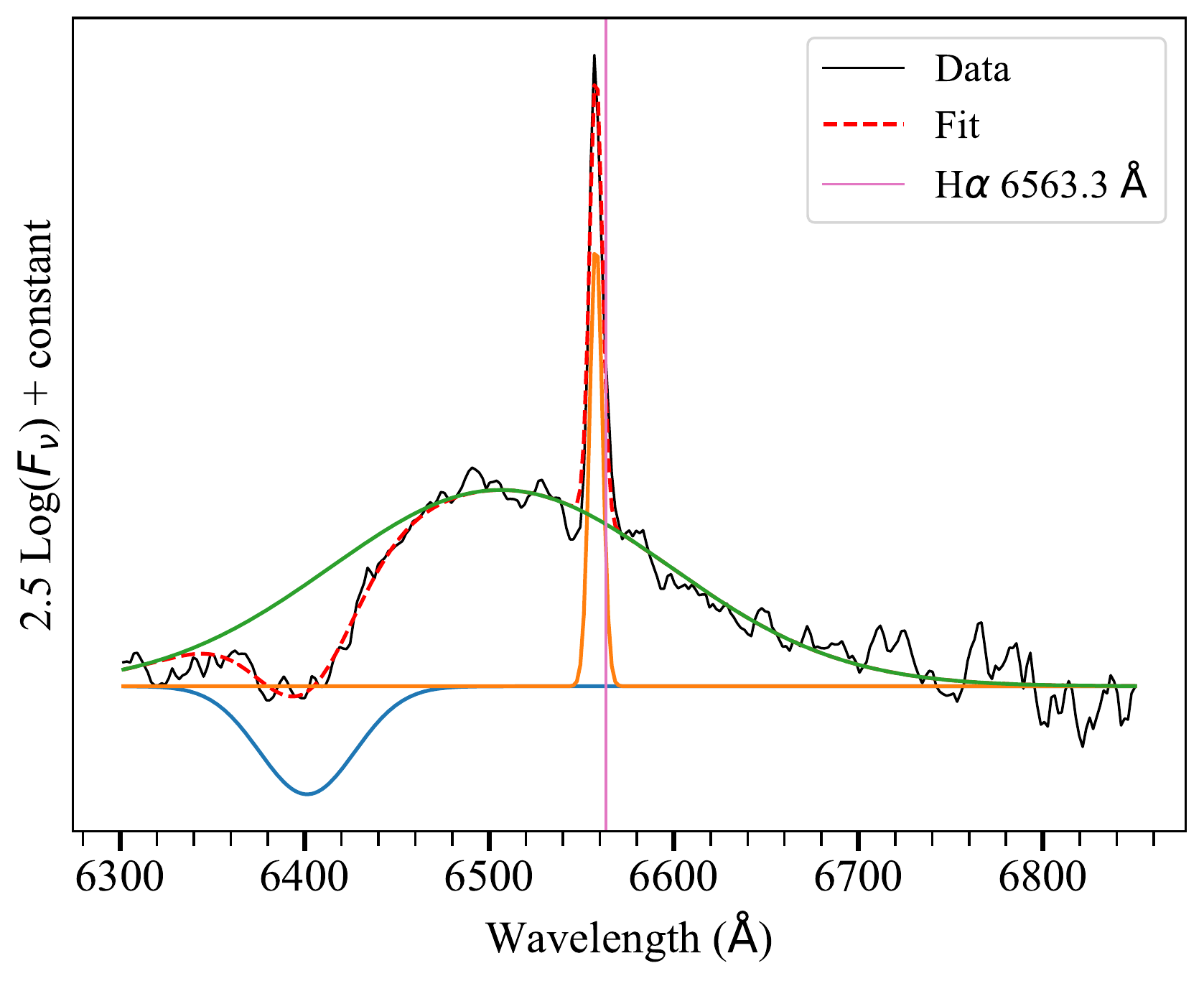}
\vspace{-1mm}
\caption{The H$\alpha$ profile for SN~2019hcc at +19~days past maximum with model profiles composed by several Gaussian profiles: a blue-shifted H$\alpha$ absorption (blue), the galaxy line (orange), a blue-shifted H$\alpha$ (green).}
\label{fig: component}
\end{figure}

In Figure~\ref{fig: component} we display a composite Gaussian function. The H$\alpha$ profile at +19~days was chosen as it is the highest resolution spectrum of SN~2019hcc, with a resolution of 6.0~\AA. As can be seen, the multi-component function provides a good fit. The fit contains an absorption and emission component to reproduce the ejecta P-Cygni profile and a narrow emission component for the host galaxy. An additional broad Gaussian component could be due to CSM interaction, however no additional component is required for the fit. The emission component %profile in the +53~days spectrum %(chosen for having the least pronounced host emission) 
was initially fitted with both a Gaussian and a Lorentzian fit, retrieving similar $\chi^2$ values. A Lorentzian profile is typically associated with scattering of photons in an optically thick CSM, and this requires a dense scattering medium \citep[e.g.][]{Reynolds_2020}. A better fit with a Lorentzian function indicates that broadening is due to electron scattering rather than expansion \citep[e.g.][]{Chatzopoulos_2011, Taddia_2013, Nicholl_2020}. However, as a Lorentzian is not a significantly better fit, this scenario is not supported.

P-Cygni theory predicts emission of Hydrogen to peak at zero rest velocity $\lambda$6563.3, however observations reveal that emission peaks are often blue-shifted \citep{Anderson_2014}. \cite{Anderson_2014} found that significant blue-shifted velocities of H$\alpha$ emission peaks are common and concluded that they are a fundamental feature of SNe spectra. This has been suggested to be due to the blocking of redshifted emission from the far side of the ejecta by an optically thick photosphere, due to a steep density profile within the ejecta \citep[e.g.][]{Reynolds_2020}. The fit allows for a blue-shifted broad emission line as well as host galaxy emission at the rest wavelength. %, as would be expected according to \cite{Anderson_2014}. 
The Doppler shift from H$\alpha$ in Figure~\ref{fig: component} is 2610$\pm$140~km/s. \cite{Anderson_2014} found blue-shifted emission velocities on the order of 2000~km/s, therefore this result is consistent. Overall, the above analysis shows that a multi-component profile is not necessary to reproduce the observed H$\alpha$ profile and hence the spectra do not show any evidence of an ejecta-CSM interaction.

%%%%%%%%%%%%%%%%%%%%%%%%%%%%%%%%%%%%%%%%%%%%%%%%%%%%%%%%%%%%%%%%%%%%%%%%%%%%%%%%

%%%%%%%%%%%%%%%%%%%%%%%%%%%%%%%%%%%%%%%%%%%%%%%%%

\section{The early `w' shaped feature: elements contribution and their nature}
\label{OII lines}

One of the most interesting features displayed by SN~2019hcc is its early `w' shaped feature resembling that of SLSNe. Understanding its nature, composition and the possibility that it is not a trademark of SLSNe~I will have important consequences during the Vera C. Rubin and the Legacy Survey for Space and Time (LSST) era. LSST will deliver hundreds of SLSNe \citep{Inserra_2021} and thousands of CC-SNe for which we might not have the luxury of multiple epoch spectroscopy.

\begin{figure}
\centering
\includegraphics[width=1\linewidth]{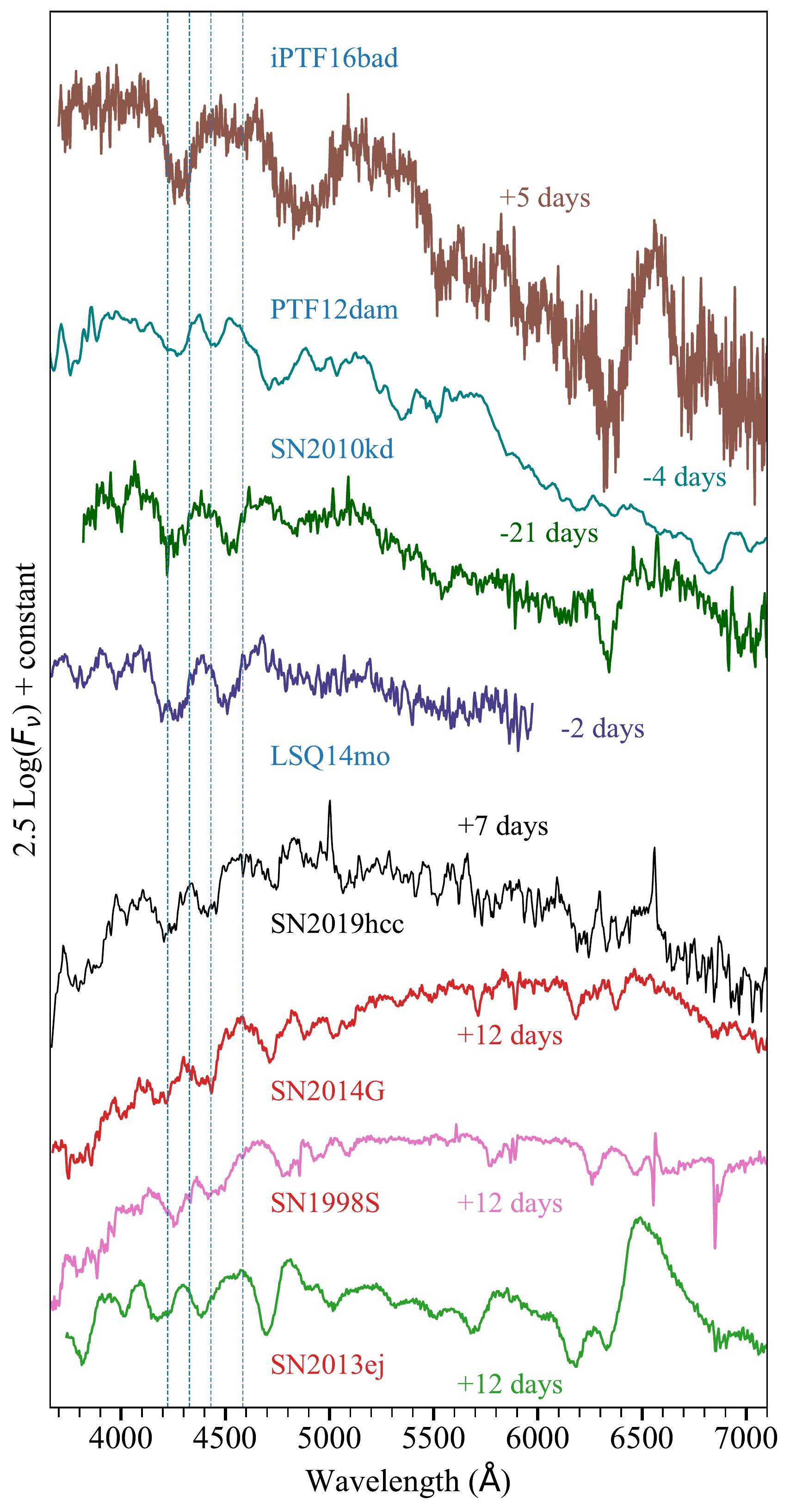}
\vspace{-1mm}
\caption{SN~2019hcc +7~days post peak is compared to moderately luminous SNe~II and SLSNe~I. These spectra are displayed in terms of F($\nu$) to emphasise the absorption features, and the wavelength is in the rest frame. SN~2014G also appears to show a `w' shaped profile at 4000-4400~\AA. The dashed lines correspond to the peaks and troughs of the {O~\sc{ii}} line region in SN~2019hcc. Text in red represents Type II, text in blue SLSN.}
\label{fig: OII_comp}
\end{figure}

Figure \ref{fig: OII_comp} shows the {O~\sc{ii}} features in the early spectrum for SN~2019hcc, together with the Type II SNe used for previous comparison, and the previous sample of SLSNe~I. The approximate location of peaks and troughs of the SN~2019hcc {O~\sc{ii}} lines are marked by dashed vertical lines for comparison. iPTF16bad \citep{halpha_2} was chosen due to the late H$\alpha$ emission, and SN~2010kd \citep{Kumar_2020} for the carbon emission which resembles H$\alpha$. PTF12dam \citep{Nicholl_2013} was chosen for being a well-sampled SLSN~I, and LSQ14mo \citep{Chen_2017c} for its similarity to SN~2019hcc with respect to the {O~\sc{ii}} feature at a similar epoch. SN~2014G, amongst the SNe~II, appears to have the strongest resemblance to SN~2019hcc, showing a similar pattern in the wavelength region around 4000~$\Angstrom$. A point to note is that SN~2019hcc does not entirely match the {O~\sc{ii}} feature in the SLSNe~I  - the redder absorption is blue-shifted in comparison.

The features usually associated with {O~\sc{ii}} are formed by many tens of overlapping lines \citep{Anderson_2018, galyam_2019b}, and can be contaminated by carbon and metal lines, and also by the presence of well-developed Balmer lines, all of which mean the features cannot be uniquely identified as {O~\sc{ii}}. Therefore, whilst SN~2019hcc, SN~2014G, and SN~1998S could be valid candidates to show {O~\sc{ii}} features as the Balmer lines are less prominent, SN~2013ej is less likely as it shows a strong H$\alpha$ profile suggesting the spectrum is dominated by H$\beta$ at $\lambda$4861 and H$\gamma$ at $\lambda$4340.

\cite{GalYam_2019} tackled the challenge of line identification with comparison of absorption lines to lists of transitions drawn from the National Institute of Standards and Technology (NIST) database. %\citep{Kramida_2018} {\bf Are you sure this is the right citation? DOes it really need a citation? NIST has been there for decades}. 
He found that {O~\sc{ii}} emission lines appear in the gaps between {O~\sc{ii}} absorption, which corresponds to the two peaks - see Figure \ref{fig: OII_comp} 2$^{\rm nd}$ and 4$^{\rm th}$ dashed lines from the left. %The entire blue region of this spectrum around 4000 $\Angstrom$ is dominated by a forest of {O~\sc{ii}} lines, and gaps in this absorption correspond to the emission features seen. 
%Manual line identification in SN spectra is complicated by the need to account for the SN expansion velocity and line blending of different elements that have very similar wavelengths. 
\cite{Anderson_2018} suggested that a change in the morphology of the spectrum in this wavelength region (between SNe) may be produced through differences in ejecta density profiles or caused by overlapping lines such as {Fe~\sc{iii}}. 

Oxygen lines appear when Oxygen is ionised by sufficiently high temperatures, 12000-15000~K \citep[e.g.][]{Extreme_SN}. However, the presence of {O~\sc{ii}} lines around 4000-4400~$\Angstrom$ might be a consequence of non-thermal excitation \citep{Mazzali_2016}. This requires a power source in the CO core of massive stars \citep{Mazzali_2016}. A lack of {O~\sc{ii}} lines would be be the product of rapid cooling or lack of non-thermal sources of excitation \citep{Quimby_2018}.

A non-thermal excitation could be in the form of strong X-ray flux from a magnetar, such as the injection of X-rays from an interaction between the SN ejecta and a magnetar wind \citep{Maeda_2007}. \cite{Metzger_2021} modelled SLSNe powered by a relativistic wind from a central engine, such as a millisecond pulsar or magnetar, which inflates a nebula of relativistic electron/positron pairs and radiation behind the expanding supernova ejecta shell. These quickly radiate their energy via synchrotron and inverse Compton (IC) processes in a broad spectrum spanning the X-ray/gamma-ray band, a portion of which heats the ejecta and powers the supernova emission. This process will be most efficient at early times after the explosion, when the column density through the ejecta is at its highest.
Non-thermal excitation could also be due to high energy electrons produced by $\gamma$-rays from the radioactive decay of $^{56}$Ni \citep{Li_2012}, however such a process would more likely be relevant at later times. It could also be produced by ejecta-CSM interaction \citep{Nymark_2006}, with a CSM rich in Oxygen producing the associated spectral features \citep{Chatz_2012}. No SLSN~I to date has shown narrow lines in its spectra \citep{Nicholl_2014, Extreme_SN}, and interaction models are yet to reproduce the observed spectra. Nevertheless, the interaction model is still favoured to reproduce the light curve evolution of some SLSNe~I \citep[e.g.][]{Chatz_2013}.

Though supposed to be typical to SLSNe~I \citep{book}, {O~\sc{ii}} lines have already been seen in other SNe, such as SN~Ibn OGLE-2012-SN-006 \citep{Pastorello_2015} and SN~Ib SN~2008D \citep{Soderberg_2008}. SN 2008D was a normal core-collapse SN with an associated X-Ray flash \citep[e.g][]{lixin_2008}, whereas OGLE-2012-SN-006 was interpreted as a core-collapse event powered by ejecta-CSM interaction \citep{Pastorello_2015}. The presence of {O~\sc{ii}} spectroscopic features here support the argument that ejecta-CSM interaction may be an important factor in maintaining the high levels of energy required to ionize Oxygen \citep{Pastorello_2015}. 

%%%%%%%%%%%%%%%%%%%%%%%%%%%%%%%%%%%%%%%%%%%%%%%%%

\subsection{Spectral Modelling}
\label{Modelling}

Reproducing the `w' shape of the first spectrum with spectral modelling could cast light on the conditions required to produce it. If the feature is reproduced by modelling Oxygen at a higher temperature than the spectra which display this feature, it would suggest non-thermal excitation is necessary to produce this feature. 

We used \textsc{Tardis} \citep{Kerzendorf_2014}, an open-source radiative transfer code for spectra modelling of SNe, to model SN~2019hcc's first spectrum. The code uses Monte-Carlo methods to obtain a self-consistent description of the plasma state and compute a synthetic spectrum. \textsc{Tardis} was originally designed for Type Ia SNe and recently improved to be used for Type II spectra \citep{Vogl_2019a}, although the time varying profile of H$\alpha$ remains difficult to reproduce. \textsc{Tardis} assumes that the ejecta is in a symmetric and homologous expansion, and as such there is a direct correlation between time since explosion and the temperature at this time.

\begin{figure}
\centering
\includegraphics[width=1\linewidth]{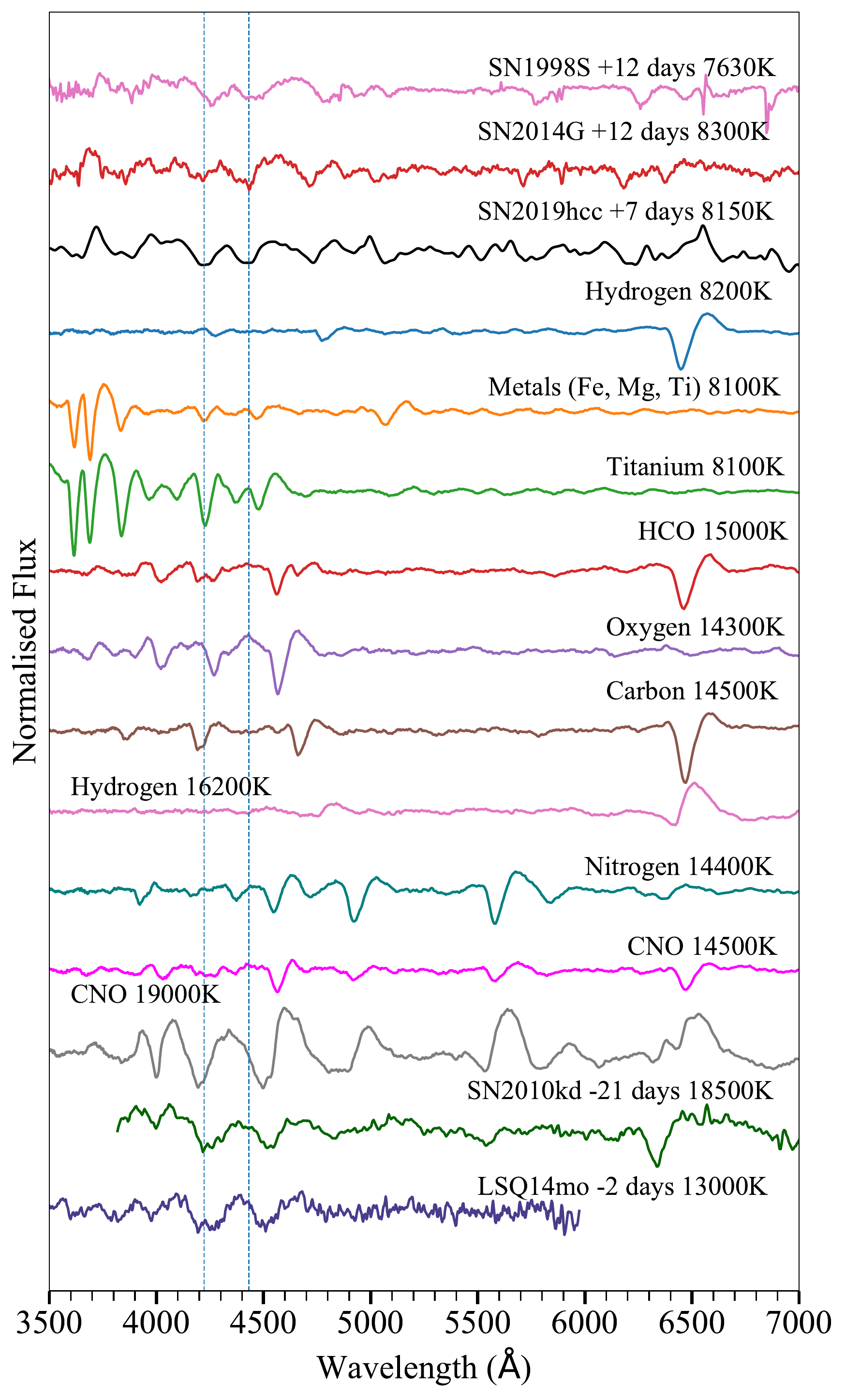}
\vspace{-1mm}
\caption{The output of the \textsc{Tardis} modelling - spectra with various abundances and temperature. The vertical dashed lines mark the absorption lines for the SN~2019hcc `w' feature.}
\label{fig: tardis}
\end{figure}

SN~2019hcc was modelled as having a uniform ejecta composition and the results are presented in Figure \ref{fig: tardis}. Model spectra were created with various abundances and temperatures and then normalised for comparison with SN~2019hcc. The temperatures were chosen to be around 8100~K (near the measured temperature of SN~2019hcc) or around 14000~K (closer to the SLSNe~I used for comparison, see Figure \ref{fig: SLSN_temp}). Higher temperatures up to around 20000~K were also considered in order to investigate the effect of the temperature on the resulting spectra. The velocity was kept constant for all spectra, at 8000~km/s (start 6000~km/s, stop 8000~km/s), similar to the photospheric velocity measured by {Fe~\sc{ii}} (see Figure~\ref{fig: velocity_comp}). Elements were investigated individually - with abundances of up to 100\% for one element. Starting from the approximate epoch and luminosity of SN~2019hcc, the spectra at approximately 8100~K were modelled by adjusting the input parameters until matching the temperature to that measured from the +7~days spectrum for SN~2019hcc after Cardelli correction, as marked in the figure. The high temperature spectra around 15000~K were found by increasing the luminosity and decreasing the time since explosion in the model.

\begin{table}
\begin{tabular}{ c c c c }
\hline 
SN name & Type & EW (blue/red) & FWHM (blue/red)\\ 
\hline 
SN~2019hcc & SN~IIL & 1.11$\pm$0.05 & 1.06$\pm$0.03 \\
SN~2014G & SN~IIL & 0.77$\pm$0.03 & 1.03$\pm$0.05\\
SN~1998S & SN~IIn & 0.94$\pm$0.06 & 0.77$\pm$0.04\\
SN~2010kd & SLSN~I & 1.39$\pm$0.07 & 1.24$\pm$0.02\\
LSQ14mo & SLSN~I & 1.61$\pm$0.06 & 1.29$\pm$0.04\\

\hline 
\end{tabular}
\caption{Equivalent widths (EW) and full with at half maximum (FWHM) of the absorption of the blue line profile over the red of the `w' feature.}
\label{tab: ew}
\end{table} 

Modelling revealed that at the lower temperature of 8100~K, Carbon, Oxygen, and Helium are not sufficiently excited to show any lines, therefore they have been omitted from the figure. However, metal (Fe, Mg, Ti) and Balmer lines do show line profiles in this region which could have the potential to reproduce the absorption lines seen for SN~2019hcc. Hydrogen does not have largely significant absorption in this region compared to these metals. Also shown in Figure \ref{fig: tardis} are elements at a higher temperature which is typical of SLSNe~I at a similar phase to SN~2019hcc's first spectrum (see Figure~\ref{fig: SLSN_temp}). These do not match well the overall spectrum of SN~2019hcc but it can be noted that Carbon, Oxygen and Nitrogen produce lines in the region of interest.

The bottom model spectrum of Figure \ref{fig: tardis} shows that at approximately 19000~K a `w' feature can be produced with a CNO composition (with an even split of abundances). Note that Nitrogen has a relatively small effect in comparison to Carbon and Oxygen in producing this shape. The `w' feature for SN~2019hcc is slightly shifted compared to the SLSNe~I used for previous comparison - such a shift is evident in the red absorption but not the blue. A possible explanation for SN~2019hcc `w' profile could be a combination of metals at a lower temperature (8100~K) and a non-thermally excited CNO layer. Considering that the temperatures of LSQ14mo and SN~2010kd are around 13000~K (at this temperature CNO does not show a `w' feature), this could confirm that these SLSNe~I require non-thermal excitation to produce this feature.

The feature of SN~1998S looks different to SN~2019hcc  - both lines of the `w'-feature have a different shape. The `w' feature in SN~1998S is likely caused by Titanium and a combination of other metals like Barium \citep{Faran_2014}, which is also seen at redder wavelengths in SN~1998S but not in SN~2019hcc. %- further discussion of 1998S is beyond the scope of this paper. 
Titanium does not look responsible for SN~2014G or SN~2019hcc as the ratios and shapes of the two profiles are different. The contribution from the combination of metals including Iron can be seen clearly in SN~2019hcc at 5169~$\Angstrom$, however Iron lines cannot account for the strong absorption in the `w' feature region. Reproducing the strength of lines would appear to require CNO abundances at higher temperatures - for example Oxygen and Carbon at approximately 14000~K could account for the broader red wing of SN~2019hcc. A combination of CNO at higher temperatures than SN~2019hcc spectrum (i.e. 8100K) and metals at 8100K could be causing the final feature. However, with the tested models it seems impossible to completely reproduce the `w' feature. Nevertheless, it appears models at T~$>14000$~K are required to reproduce the strength of the absorption, suggesting a non-thermal excitation responsible for the CNO elements SN~2019hcc at +7~days. 

Equivalent width (EW) ratios are measured in order to provide a more quantitative analysis of the feature. These are reported in Table \ref{tab: ew} in the form of the EW of the blue line over the red one, as well as the same ratio for full width at half maximum (FWHM). Of the SNe~II, only SN~2019hcc has an EW over 1. The SLSNe~I in this table also have a ratio over 1 and are larger with respect to that of SN~2019hcc. In both cases the SLSNe~I have a slightly higher FWHM than the SNe~II although this is not statistically conclusive due to the small size of the sample. These ratios cannot offer anything conclusive as it suggests all these `w' features are of a slightly different nature, and could possibly be affected by temperatures, abundances, non-thermal excitation, or the presence of other lines such as metal lines. Possibly SN~2014G could also be non-thermally excited, or have different metal contributions, though its nature looks different to the other SNe as it is the only spectrum with a significantly stronger red line than blue.

In summary, at temperatures of approximately 19000~K CNO could reproduce the `w' feature. Some absorption in this region at a temperature of 8100~K could be caused by metal lines e.g. Titanium, however this cannot entirely account for the `w' feature in SN~2019hcc spectrum. Metals would also produce stronger lines at bluer wavelengths (3500-4000~$\Angstrom$) which are not seen in SN~2019hcc, though these could be obscured by yet more lines in this region. For thermally exciting CNO much higher temperatures are needed than that observed for SN~2019hcc, therefore non-thermal excitation may be required to produce such features in SN~2019hcc. This appears to also be the case for LSQ14mo and SN2010kd, which show the feature despite LSQ14mo being almost 6000~K short of the required excitation temperature.

{He~\sc{i}} can also be non-thermally excited, however this excitation usually comes from CSM interaction at the outer boundary of the ejecta \citep[e.g.][]{Chevalier_1994}, whereas for the non-thermal excitation of {O~\sc{ii}} in this scenario the exciting X-ray photons would originate from the central engine. The ejecta Helium region would be further away than the Oxygen region for these central high-energy photons which, in our proposed scenario, would explain the absence of {He~\sc{i}} in the first spectrum of SN~2019hcc. Additionally, though the abundance of Oxygen in the progenitor is relatively low compared to other elements such as Hydrogen, the first spectrum is relatively featureless so {O~\sc{ii}} is not competing with other lines in this region.

Hence, the next question to address is what could cause the non-thermal excitation of such CNO lines.

%%%%%%%%%%%%%%%%%%%%%%%%%%%%%%%%%%%%%%%%%%%%%%%%%%%%%%
\subsection{Ejecta-CSM interaction scenario}

The presence of {O~\sc{ii}} lines could be the consequence of ejecta-CSM interaction \citep[e.g.][]{Pastorello_2015}. \cite{Mazzali_2016} suggested that X-rays would be required for the non-thermal excitation of {O~\sc{ii}} lines, and these X-rays could originate from interaction \citep{Nymark_2006}. However, \cite{Chevalier_1994} suggested that in ejecta-CSM interaction with a SN density profile consistent with that of an RSG progenitor, as with the majority of Type II, the photons produced would be primarily in the UV-range, thus not providing sufficient non-thermal excitation to ionise the Oxygen. %Ejecta-CSM interaction could also be used to explain the H$\alpha$ seen in the spectra of SN~2019hcc as collision with an outer shell ejected prior to explosion, as seen in e.g iPTF13ehe \citep{halpha_1}.

There are no distinctive narrow emission lines in the spectrum of SN~2019hcc, nor is there any unusual behaviour in the light curve such as multiple peaks or undulations which would suggest collision with a shell \citep[e.g.][]{Nicholl_2016,Inserra_2017}. A possible HV component of H$\alpha$ blue-ward of the main emission could be indicative of early weak/moderate CSM-ejecta interaction - as this interaction may excite the Hydrogen to cause a second, high-velocity absorption feature \citep[e.g.][[]{SNIIbook}. However, our results on the HV H$\alpha$ analysis reported in Section~\ref{sec:hv} suggest that the presence of a HV H$\alpha$ is unlikely with the absorption blue-ward than H$\alpha$ plausibly associated with {Si~\sc{ii}}. The overall H$\alpha$ profile was also analysed and decomposed in multiple components investigating the nature of the profile. However, it was found that no additional components are required to reproduce the shape aside from the expected ejecta P-Cygni and the narrow H$\alpha$ line from the host galaxy. Therefore, CSM-ejecta interaction is not a viable source for generating high-energy photons capable of non-thermally excited the {O~\sc{ii}} lines in SN~2019hcc. %The scenario of a collision with a previously ejected hydrogen shell can also be discounted as a result.

%%%%%%%%%%%%%%%%%%%%%%%%%%%%%%%%%%%%%%%%%%%%%%%%%%%%%%%%%
\subsection{Magnetar scenario}

A magnetar could produce the non-thermal excitation required to ionize Oxygen and produce the {O~\sc{ii}} features \citep[e.g.][]{Mazzali_2016}. \cite{Dessart_2012} suggested the magnetar's extra energy heats material and thermally excites the gas. Alternatively, \cite{Soker_2016} and \cite{Soker_2017} suggested that magnetar-driven SLSNe are powered not by the neutrino-driven mechanism but a jet feedback mechanism from jets launched at magnetar birth. These high energy jets could potentially provide the energy to drive {O~\sc{ii}} excitation at early times, and have been used to link magnetars to Gamma Rays Bursts (GRBs) \citep{Wheeler_2000}. The generation of a non-relativistic jet during the early supernova phase is a consequence in both the core-collapse and magnetar models of GRBs \citep{Burrows_2007}.

\cite{Kasen_2010} suggested that a magnetar birth is likely to happen in a few percent of all core-collapse supernovae, and may naturally explain some of the brightest events seen. \cite{magnetar} found that magnetar-powered models can actually generate a diversity of Hydrogen-rich SNe, both ordinary and brighter ones. Through their modelling, it was found that the observational appearance of SNe~II powered by magnetars can be extremely varied and can also mimic those of normal SNe~IIP. Magnetars are thought to form by fast rotation in the collapsing Iron core \citep{Duncan_Thompson_1992}. It is suggested that magnetars are preferentially formed in the most massive stars collapsing to a neutron star - with a progenitor mass in excess of 40M$_{\odot}$ \citep{Davies_2009}. However, it has also been suggested that magnetars do not require massive progenitors to form - alternatives could be a `fossil-field' model, where a seed B-field is inherited from the natal molecular cloud \citep{Davies_2009} or an interacting binary system which causes spin-up in the collapsing CO-core \citep{Cantiello_2007}.

\cite{Chen_2017} found an apparent correlation between magnetar spin-down period and host metallicity from a sample of 19 SLSNe~I, indicating that faster-rotating magnetars reside in more metal-poor environments. 
%Note however that \cite{Nicholl_2017} found no correlations between any model parameters and the properties of SLSN host galaxies for a larger sample of 38 SLSNe~I. 
Such a correlation could be a consequence of several factors - \cite{Martayan_2007} found that massive stars rotate more rapidly at lower metallicity (0.2~$Z/Z_{\odot}$) than solar, whilst \cite{Mokiem_2007} found in low metallicity environments mass loss of rotating stars is reduced. However, the spin periods of low metallicity stars and neutron stars would also very likely be affected by other parameters. Generally, the greater the spin period, the greater the peak luminosity \citep{Kasen_2010,Inserra_2013}, therefore a high metallicity host environment could be correlated with low luminosity explosions powered or affected by a magnetar.

\begin{figure}
\centering
\includegraphics[width=1\linewidth]{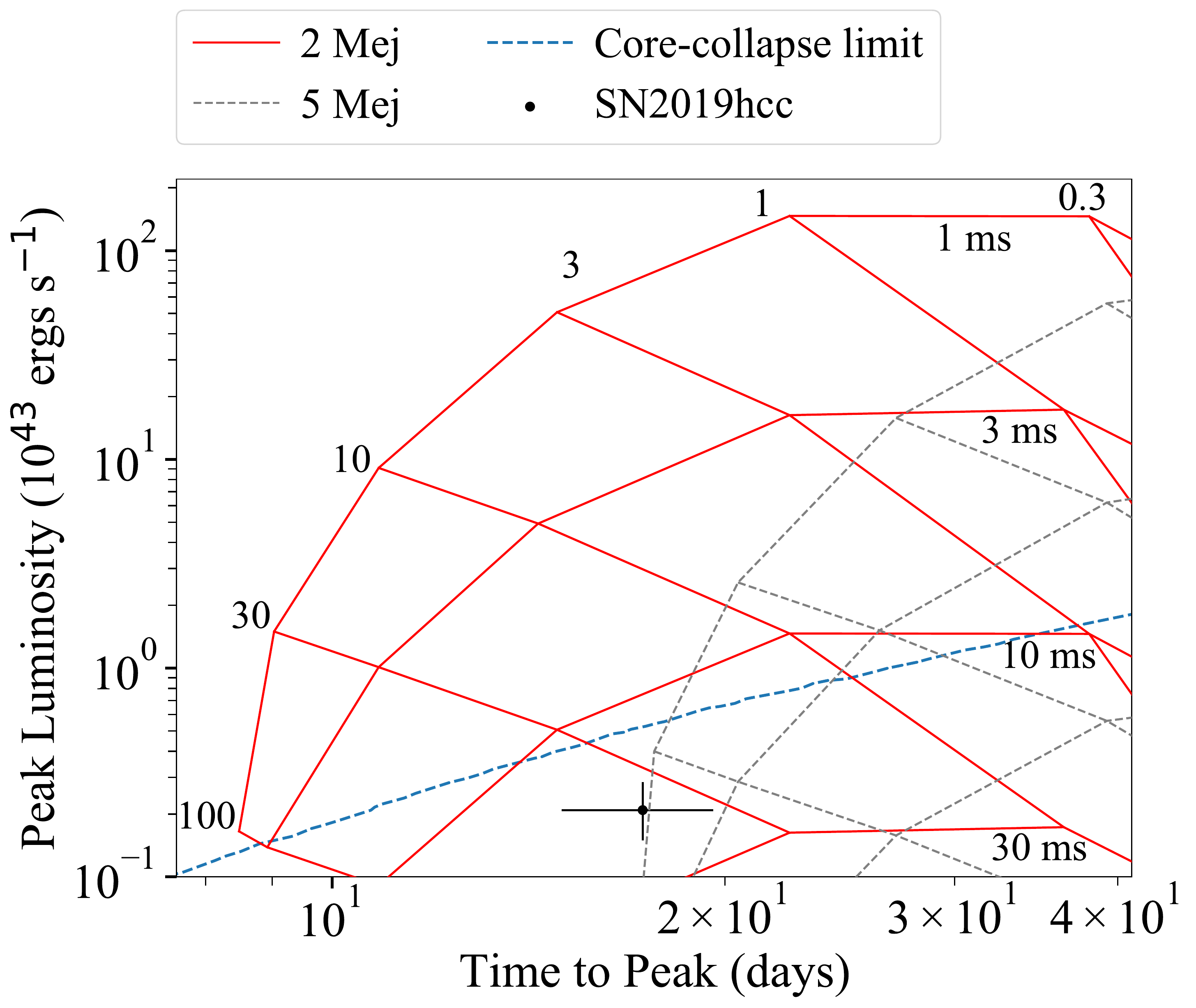}
\vspace{-1mm}
\caption{The effect of ejecta mass on a magnetar model grid (see text for further information about the grid limits and model used) - with an opacity of 0.34~g~cm$^{-3}$ suited for a H-rich ejecta. The grid markers are for Pms in ms and B14 in B/10$^{14}$~G.}
\label{fig: magnetar_35}
\end{figure}

From the equations in \cite{Kasen_2010}, a grid of B14 (B/10$^{14}$~G) and Pms (the spin period in ms) of a magnetar as a function of the peak luminosity and rise time was produced, using the code presented in \cite{Inserra_2013}. Multiple grids were created by varying the ejecta mass in the model, in order to investigate its effect. Figure~\ref{fig: magnetar_35} shows an ejecta mass of 2~M$_\odot$ vs. 5~M$_\odot$. These ejecta masses were chosen based on the bolometric light curve fitting of SN~2019hcc \citep[using the code of][]{Inserra_2013} and that of SN~2014G which is one of the other potential Type II showing the `w'-shaped feature. We retrieved an ejecta mass of approx. 2.3~M$_{\odot}$ and 5.0~M$_{\odot}$, respectively. The fitting was focused on matching the rise time and peak magnitude rather than attempting to accurately reproduce the entire shape of the light curve including the tail, as this is also affected by other factors such as $^{56}$Ni or CSM interaction. The range of values in the grid are based on the fact that the neutron stars cannot spin faster than 1~ms without breaking up and that spin periods $<$30~ms can substantially modify the thermal evolution of the supernova \citep{Kasen_2010}, while B values are those retrieved from galactic magnetars $\approx 10^{14} - 10^{15}$~G \citep{Woods_2006}. This figure shows that increasing the ejecta mass, but preserving B14 and Pms, would result in a longer rise time with the luminosity not as significantly affected. SN~2019hcc's location in this parameter space (see Figure~\ref{fig: magnetar_35}) shows that a lower luminosity supernova (i.e. not a SLSN) could be produced by a high magnetic field and a relatively lower spin. The blue dashed line represents the core-collapse limit for peak luminosity vs. rise time \citep{Extreme_SN}. \cite{Sukhbold_2017} also presented a proof-of-concept model of a magnetar mechanism producing Type IIP light curve properties for a range of initial spin periods and equivalent dipole magnetic field strengths, and found for a SNe of peak bolometric luminosity of $\sim$42.5, approximately that of SN~2019hcc, one would expect a Pms of 2ms and a B14 of 100 - this agrees very well with the 5~M$_{\odot}$ model in Figure~\ref{fig: magnetar_35}.

This modelling suggests it is possible to have a magnetar formed as a remnant without injecting further substantial energy to the supernova event leading to superluminous brightness. This could provide sufficient non-thermal contribution to excite the {O~\sc{ii}} lines which appear in the early spectra. The sub-solar metallicity found in Section \ref{Observations} would not provide support for the tentative hypothesis of a correlation between host environment metallicity and magnetar luminosity, as the metallicity is similar to that of the the typical low metallicity environments of SLSNe~I, whilst the luminosity is typically lower than that of SLSNe~I.

%\cite{Chen_2013} - magnetar progenitor for SLSN I likely Wolf-Rayet or Carbon-Oxygen star as no sign of hydrogen

%\cite{Milisavljevic_2018} found that 6.2 years after explosion forbidden transition emission lines of oxygen and sulfur were detected with expansion velocities of ≈ 2300~Km s−1 in SNe suspected to be magnetar powered. They predicted SNe harboring influential pulsar/magnetar wind nebulae will evolve into a late-time phase dominated by forbidden oxygen transitions. Unfortunately there are no late time spectra for SN~2019hcc. 

%%%%%%%%%%%%%%%%%%%%%%%%%%%%%%%%%%%%%%%%%%%%%%%%%%%%%%

%\subsubsection{Convective Overshoot}

%Convective overshoot and resultant chemical mixing affects the stellar structure, surface abundances and effective temperature of massive stars. Convective flow penetrates the stably stratified region beyond, leading to mixing \cite{convective}.

%The external convection zone (when present) dredges out the CNO products from the hydrogen shell burning layers to the surface - therefore this may affect the surface abundance of these elements.

%\cite{Mazzali_2016} found that greater abundance of oxygen will not produce stronger lines where there is oxygen already present, which would be the case for a massive core-collapse star. However the high velocity of the {O~\sc{ii}} lines measured could suggest that oxygen is higher in the ejecta rather than closer to the core as would be expected \cite{Quimby_2018}, therefore mixing may have brought it closer to the surface.

%%%%%%%%%%%%%%%%%%%%%%%%%%%%%%%%%%%%%%%%%%%%%%%%%%%%%%%%%%%%%%%%%%%%%%%%%%%%%%%%
\section{Conclusions}
\label{Conclusion}

%The {O~\sc{ii}} lines were investigated through comparison with other SNe II and SLSNe I, and through modelling with TARDIS. The first spectrum of SN~2019hcc appears relatively featureless aside from a `w' feature around 4000 $\Angstrom$, which is similar to that observed in SLSNe I, though the redder absorption appears to be relatively blueshifted. The spectrum of SN~2014G is also similar to SN~2019hcc, both with respect to the `w' shape and other features such as {Fe~\sc{ii}} and the Balmer lines. Modelling with TARDIS suggested this feature could be reproduced by a CNO layer heated to 19000~K for the bluer absorption, and by metal and Balmer lines at 7000~K for the redder absorption. In order for these to coexist non-thermal excitation would be required to produce the excited CNO. The discussion which follows will consider the means of non-thermal excitation either by a magnetar or interaction, and any signatures of these mechanisms in SN~2019hcc.

The first spectrum of SN~2019hcc appears relatively featureless aside from a `w' feature around 4000 $\Angstrom$, characteristic of {O~\sc{ii}} lines typical of SLSNe~I. The redder absorption appears to be relatively blue-shifted with respect to SLSNe~I. The spectra show a clear H$\alpha$ profile from +19~days, as well as spectral similarity to various literature SNe~II, and the bolometric light curve evolution is that of a SNe~IIL. The host metallicity was sub-solar, a value lower than the typical Type II SNe \citep{Gutierrez_2018}. The temperature and colour evolution were typical of a Type II.

Such a `w'-shaped feature (usually and historically) attributed to {O~\sc{ii}} has never been identified and analysed in SNe~II as such and only recognised in SN~2014G thanks to the analysis reported in this paper. 
Modelling of this `w' feature using \textsc{Tardis} \citep{Kerzendorf_2014} suggested it could be produced by the excitation of CNO at a temperature of 19000~K, which is more than twice that measured from the spectrum, suggesting these lines would therefore be non-thermally excited. Another result of the modelling was that absorptions at these wavelengths could also be the result of metal lines at 8100~K, a temperature in agreement with that measured. 
In SLSNe~I these lines have been suggested as excited by X-rays produced by a magnetar, or alternatively CSM-ejecta interaction. As there is a lack of any sign of interaction both in the light curve and spectra, aside from a tentative HV component, and potential interaction at late epochs, the CSM-ejecta interaction at early time is disfavoured. We built a model grid, following the work of \cite{Kasen_2010} and using the code by \cite{Inserra_2013}, and found that a magnetar could be formed as a remnant in a Type II. This would require that the magnetar does not provide enough additional energy to the supernova event to power up the light curve to superluminous luminosities. The magnetar remnant could therefore non-thermally excite the Oxygen whilst not having a significant contribution to the light curve evolution.
Therefore, combining such results with those of the spectral modelling, we conclude that the `w' feature seen in SN~2019hcc's first spectrum could be due to a combination of non-thermally excited CNO and thermally excited metal lines.

The object here presented could then bridge the gap between SLSNe~I and normal luminosity core collapse supernovae, as well as reveal more about magnetar formation requirements and mechanisms. Our analysis also shows that a magnetar is a viable remnant of a Type II supernova explosion, the effects on which could be observed in the form of an early `w'-shaped profile around 4000--4400~\AA. This would suggest that such lines are not exclusive to SLSNe~I and cannot be used as a sole feature to classify those extreme transients.

%However, {O~\sc{ii}} lines have not been seen in SNe~II, and require additional non-thermal excitation \citep{Mazzali_2016}. 

\section*{Acknowledgements}
Based on observations collected at the
European Organisation for Astronomical Research in the Southern Hemisphere, Chile, as part of ePESSTO/ePESSTO+ (the extended Public ESO Spectroscopic Survey for Transient Objects Survey).
ePESSTO+ observations were obtained under ESO program ID 1103.D-0328 (PI: Inserra).

EP would like to thank Stuart Sim for useful discussion on the working of \textsc{Tardis}. This research made use of \textsc{Tardis}, a community-developed software
package for spectral synthesis in supernovae
\citep{Kerzendorf_2014, kerzendorf_wolfgang_2019}.
The development of \textsc{Tardis} received support from the Google Summer of Code initiative and from ESA's Summer of Code in Space program. \textsc{Tardis} makes extensive use of Astropy and PyNE.

TWC acknowledges the funding provided by the Alexander von Humboldt Foundation and the EU Funding under Marie Sk\l{}odowska-Curie grant H2020-MSCA-IF-2018-842471.
TMB was funded by the CONICYT PFCHA / DOCTORADOBECAS CHILE/2017-72180113.
MG is supported by the Polish NCN MAESTRO grant 2014/14/A/ST9/00121.
MN is supported by a Royal Astronomical Society Research Fellowship.
ACK: L.G. was funded by the European Union's Horizon 2020 research and innovation programme under the Marie Sk\l{}odowska-Curie grant agreement No. 839090. This work has been partially supported by the Spanish grant PGC2018-095317-B-C21 within the European Funds for Regional Development (FEDER).
G.L. is supported by a research grant (19054) from VILLUM FONDEN

This work makes use of observations from the Las Cumbres Observatory global telescope network.  The LCO team is supported by NSF grants AST-1911225, AST-1911151, and NASA grant 80NSSC19K1639.
This paper is also based on observations made with Swift (UVOT) and the Liverpool Telescope (LT). The Liverpool Telescope is operated on the island of La Palma by Liverpool John Moores University in the Spanish Observatorio del Roque de los Muchachos of the Instituto de Astrofisica de Canarias with financial support from the UK Science and Technology Facilities Council. LCO data have been obtained via OPTICON proposals (IDs: SUPA2020B-002 OPTICON 20B/003 and SUPA2019B-007 OPTICON 19B-009). The OPTICON project has received funding from the European Union's Horizon 2020 research and innovation programme under grant agreement No 730890.
This work has made use of data from the Asteroid Terrestrial-impact Last Alert System (ATLAS) project. The ATLAS project is primarily funded to search for near earth asteroids through NASA grants NN12AR55G, 80NSSC18K0284, and 80NSSC18K1575; byproducts of the NEO search include images and catalogs from the survey area. This work was partially funded by Kepler/K2 grant J1944/80NSSC19K0112 and HST GO-15889, and STFC grants ST/T000198/1 and ST/S006109/1. The ATLAS science products have been made possible through the contributions of the University of Hawaii Institute for Astronomy, the Queen’s University Belfast, the Space Telescope Science Institute, the South African Astronomical Observatory, and The Millennium Institute of Astrophysics (MAS), Chile.
This work has made use of data from the European Space Agency (ESA) mission {\it Gaia} (\url{https://www.cosmos.esa.int/gaia}), processed by the {\it Gaia} Data Processing and Analysis Consortium (DPAC, \url{https://www.cosmos.esa.int/web/gaia/dpac/consortium}). Funding for the DPAC has been provided by national institutions, in particular the institutions participating in the {\it Gaia} Multilateral Agreement.
Based on observations obtained at the Southern Astrophysical Research (SOAR) telescope, which is a joint project of the Minist\'{e}rio da Ci\^{e}ncia, Tecnologia e Inova\c{c}\~{o}es (MCTI/LNA) do Brasil, the US National Science Foundation`s NOIRLab, the University of North Carolina at Chapel Hill (UNC), and Michigan State University (MSU).
This work has made use of data from the GROND instrument at the 2.2 MPE telescope at La Silla, Chile. Part of the funding for GROND (both hardware as well as personnel) was generously granted from the Leibniz-Prize to Prof. G. Hasinger (DFG grant HA 1850/28-1). GROND data were obtained under ESO programme ID 0103.A-9099.
This research has made use of the NASA/IPAC Extragalactic Database (NED), which is funded by the National Aeronautics and Space Administration and operated by the California Institute of Technology.

This research made use of Photutils, an Astropy package for detection and photometry of astronomical sources \citep{Bradley_2020}. 
Based on data products from observations made with ESO Telescopes at the La Silla or Paranal Observatories under ESO programme ID 179.A-2010. 

\textsc{iraf} is distributed by the National Optical Astronomy Observatories, which is operated by the Association of Universities for Research in Astronomy, Inc. (AURA) under cooperative agreement with the National Science Foundation.
This research made use of NumPy \citep{Harris_2020}, Matplotlib \citep{Hunter_2007} and Astropy \citep{astropy_2013,astropy_2018}.

\section*{Data Availability}
The data underlying this article are available in the article and in its online supplementary material. The photometry code is available at the following GitHub account https://github.com/eparrag1/Photometry.

%\section{Source Code}
%\label{Code}
%The code used for the photometry can be found at https://github.com/eparrag1/Photometry

%%%%%%%%%%%%%%%%%%%% REFERENCES %%%%%%%%%%%%%%%%%%

\bibliographystyle{mnras}
\bibliography{eparrag}

%%%%%%%%%%%%%%%%%%%%%%%%%%%%%%%%%%%%%%%%%%%%%%%%%%

%%%%%%%%%%%%%%%%% APPENDICES %%%%%%%%%%%%%%%%%%%%%

\newpage
\appendix

\section{Photometry Code}
\label{Appendix: Code}
The pipeline of the photometry code which was used to produce the light curve from the LCO and LT photometry is here described. The total flux is calculated from the pre-reduced data (bias, flat) by the Iteratively Subtracted PSF Photometry from PhotUtils \citep{Bradley_2020} - the Point Spread Function (PSF) is taken to be a Gaussian as this is found to produce a good fit and is an acceptable approximation. The PSF fitting is confined to a 50-pixel-width square around the central SN coordinates. With an average FWHM below 10 pixels, this size is assumed to safely include all the associated flux. Alternatives to the Gaussian PSF were also considered - such as an ePSF (effective PSF) constructed from reference stars, as well as %experimentation with using a measured FWHM to selected the parameters for the fitting. 
aperture photometry. The Gaussian PSF was found to be the method where the scatter between adjacent points was minimised. The magnitude is calculated as below (where $ZP$ is the zero-point):

\begin{equation}
Mag = ZP - 2.5 \, log(Counts/Exposure)
\label{eqn 4}
\end{equation}

%no additional sources within a closer region of approx. 5 pixels, and no more than 5 sources fitted in the full 50-pixel area
Valid PSF fits are filtered by setting a threshold of 3$\sigma$ and requiring no close stars which would suggest an unreliable fit. These constraints are optimised through variation and inspection of residuals. The uncertainty is obtained by combining in quadrature the uncertainty in the fit given by the PSF and the uncertainty in the image. The uncertainty in the image is given by:

\begin{equation}
Error = \sqrt{\frac{Counts+Sky}{Gain} + Npix\times(Readnoise+Sky)}
\label{eqn 5}
\end{equation}

Where $Sky$ is the sky counts over an area the size of the SN, calculated by finding the sigma-clipped mean in the environment surrounding the SN and multiplying by the number of pixels, $Npix$ in the above.  $Gain$ and $Readnoise$ come from the header of each fits file. The equation below shows how this uncertainty in counts is converted to magnitude.

\begin{equation}
Error = \frac{2.5}{ln(10)}  \sqrt{\frac{Error(Count)}{Count(Total)}}
\label{eqn 6}
\end{equation}

This uncertainty is then combined in quadrature with the extinction and the colour uncertainties, which are taken as 0.03 and 0.011 respectively. These values are taken from \cite{Valenti_2016} for one telescope and is carried over as an approximation for the others. Such an assumption might appear unreasonable, but it is indeed acceptable as these terms are a small contribution to the uncertainy budget, and these values are roughly representative (ranges are 0.02-0.09 for extinction, and 0.011-0.036 for the colour). Cosmic ray artifacts are removed using \textsc{lacosmic} \citep{lacosmic}.

The $ZP$ are found by fitting the PSF to reference stars and reversing the magnitude calculation. This is achieved in the code by accessing the Pan-STARRS \citep{Chambers_2016} catalogue using Vizier \citep{Vizier} and selecting all available stars in a 5 arcmin radius around the SN coordinates (see Figure~\ref{fig: finder}). To improve the quality of the PSF fit, multiple images taken on the same night (when available) were (and can be in a general workflow) aligned and stacked using the SNOoPY (SuperNOva PhotometrY) package \footnote{SNOoPy is a package for SN photometry using PSF fitting and/or template subtraction developped by E. Cappellaro. A package description can be found at http://sngroup.oapd.inaf.it/ecsnoopy.html.}. Template subtraction for this code is as follows: host images (which could be combined using SNOoPY, excluding poorer images) and the flux of both the host image and each SN image are found using the PSF fitting method described above. Equation \ref{eqn 4} is used to convert the host flux to what it would be if it had the same $ZP$ and exposure of the SN image, then the fluxes are subtracted, and the uncertainties propagated.

\begin{figure}
\centering
\includegraphics[width=1\linewidth]{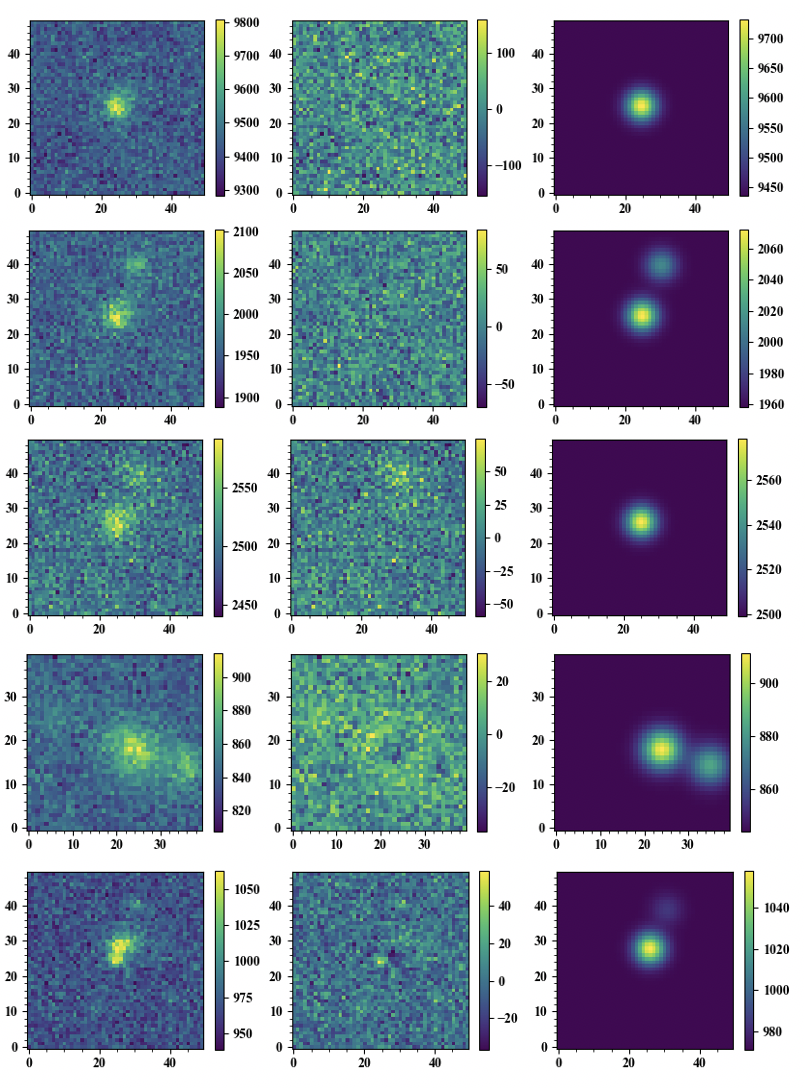}
\vspace{-1mm}
\caption{Demonstration of PSF fitting for sample image data. Column 1: Image Data. Column 2: Residual, Column 3: PSF fit. Row 1: $i-$band MJD=58655, Row 2: $z-$band MJD=58674, Row 3: $z-$band MJD=58684, Row 4: Host galaxy in $i-$band, Row 5: $r-$band MJD=58703. All images above were taken by the Liverpool Telescope \citep{Steele_2004}. The images display the 50-by-50 pixel width square centred on the coordinates of SN~2019hcc, and the colourbar displays the counts as represented by the intensity.}
\label{fig: code_fig}
\end{figure}

Figure \ref{fig: code_fig} displays the PSF fitting for a few example images. The first column displays the image data, whilst the second and third show the residual and PSF fit, respectively. As can be seen, the Gaussian PSF fit can produce relatively clean residual images, and the code recognises multiple sources. 

%Code_fig.png

\section{Data}
\label{Appendix: Data}

\begin{table*}
\label{Resolution_table}
\caption{Spectroscopy Data as displayed in Figure~\ref{fig: spectra}. The resolutions of the spectra are found from measuring the skylines using \textsc{iraf}, excluding the SOAR spectrum resolution which was was taken from http://www.ctio.noao.edu/soar/content/goodman-spectrograph-gratings.} %Source: The resolution could not be measured for the SOAR instrument and was taken from http://www.ctio.noao.edu/soar/content/goodman-spectrograph-gratings}
\begin{tabular}{c c c c c c}
\hline
Epoch &  Phase from maximum (days) & Instrument & Grisms & Range ($\Angstrom$) & Resolution ($\Angstrom$) \\ 
\hline
58643 & 7  & EFOSC2 & Gr 11 & 3380-7520 & 13.7\\ 
58655 & 19  & SOAR & 400mm & 3200-8500  & 6.0\\  
58665 &  29  & EFOSC2 & Gr 13 & 3685-9315 & 25.7\\  
58689 &  53  & EFOSC2 & Gr 13 & 3685-9315 & 17.4\\
58717 &  81  & EFOSC2 & Gr 13 & 3685-9315 & 17.1\\
58814 &  178  & EFOSC2 & Gr 13 & 3685-9315 & 17.3 \\
\hline
59149 &  Host Spectrum & EFOSC2 & Gr 13 & 3685-9315 & 16.0 \\
\hline

\end{tabular}
\end{table*}

\begin{table*}
\caption{The measured apparent magnitudes of the host galaxy for SN~2019hcc from LT and LCO images.}
\begin{tabular}{c | c | c}
\hline
Filter &  Apparent Magnitude\\ 
\hline
B &	20.70 (0.18) \\
V &	20.05 (0.23)\\
g & 20.68 (0.24)\\
r & 20.67 (0.28)\\
i & 20.37 (0.29) \\
z & 20.56 (0.18)\\
\hline

\end{tabular}
\end{table*}

\begin{table*}
\caption{NIR GROND magnitudes as seen in Figure~\ref{fig: Photometry}}
\begin{tabular}{c|c|c|c|c}
\hline
MJD	&	Phase from maximum (days)	&	J	&	H	&	K	\\ 
\hline
58644	&	8	&	18.08 (0.17)	&	17.87 (0.25)	&	17.64 (0.50)	\\
58648	&	12	&	18.08 (0.17)	&	17.90 (0.25)	&	17.80 (0.03)	\\
58661	&	25	&	18.23 (0.18)	&	17.64 (0.23)	&	17.12 (0.28)	\\
58667	&	31	&	18.26 (0.18)	&	17.91 (0.23)	&	17.56 (0.32)	\\
58674	&	38	&	18.33 (0.17)	&	18.00 (0.23)	&	-	\\
\hline

\end{tabular}
\end{table*}

\begin{table*}
\caption{Photometry data shown in Figure~\ref{fig: Photometry}.}
\begin{tabular}{c|c|c|c|c|c|c|c|c}
\hline
MJD	&	Phase from maximum (days)	&	B	&	V	&	g	&	r	&	i	&	z	&	Telescope \\
\hline
58644	&	8	&	-	&	-	&	18.84 (0.11)	&	18.73 (0.09)	&	18.78 (0.11)	&	18.69 (0.12)	&	GROND	\\
58647	&	11 	&	-	&	-	&	19.19 (0.31)	&	19.01 (0.22)	&	19.51 (0.26)	&	19.22 (0.26)	&	LT	\\
58647	&	11 	&	-	&	-	&	-	&	-	&	19.37 (0.29)	&	-	&	LCO	\\
58648	&	12 	&	-	&	-	&	19.15 (0.03)	&	18.87 (0.10)	&	18.90 (0.12)	&	18.75 (0.12)	&	GROND	\\
58651	&	15 	&	-	&	-	&	-	&	-	&	19.67 (0.44)	&	-	&	LCO	\\
58653	&	17 	&	-	&	-	&	-	&	-	&	19.48 (0.40)	&	19.23 (0.38)	&	LCO	\\
58653	&	17 	&	-	&	-	&	-	&	-	&	19.06 (0.42)	&	-	&	LT	\\
58656	&	20 	&	-	&	-	&	19.50 (0.40)	&	19.23 (0.38)	&	19.46 (0.32)	&	19.59 (0.45)	&	LT	\\
58657	&	21 	&	20.30 (0.30)	&	19.46 (0.36)	&	19.65 (0.36)	&	19.44 (0.35)	&	19.78 (0.36)	&	-	&	LCO	\\
58659	&	23 	&	19.48 (0.29)	&	19.58 (0.37)	&	19.75 (0.37)	&	19.43 (0.29)	&	19.80 (0.36)	&	-	&	LCO	\\
58660	&	24 	&	20.60 (0.31)	&	19.49 (0.34)	&	19.96 (0.39)	&	19.47 (0.27)	&	19.77 (0.33)	&	-	&	LCO	\\
58661	&	25 	&	-	&	-	&	19.72 (0.03)	&	19.16 (0.17)	&	19.22 (0.17)	&	19.14 (0.17)	&	GROND	\\
58662	&	26 	&	20.52 (0.25)	&	19.62 (0.36)	&	19.86 (0.37)	&	19.47 (0.29)	&	-	&	-	&	LCO	\\
58662	&	26 	&	-	&	-	&	20.04 (0.42)	&	19.46 (0.25)	&	19.84 (0.26)	&	19.63 (0.29)	&	LT	\\
58664	&	28 	&	20.86 (0.34)	&	19.69 (0.38)	&	20.17 (0.43)	&	19.72 (0.31)	&	-	&	-	&	LCO	\\
58665	&	29 	&	20.70 (0.29)	&	19.74 (0.39)	&	20.03 (0.40)	&	19.51 (0.29)	&	19.88 (0.36)	&	-	&	LCO	\\
58667	&	31 	&	-	&	-	&	19.83 (0.03)	&	19.34 (0.03)	&	19.28 (0.14)	&	19.23 (0.14)	&	GROND	\\
58670	&	34 	&	-	&	-	&	20.46 (0.51)	&	19.63 (0.27)	&	19.79 (0.26)	&	19.67 (0.32)	&	LT	\\
58673	&	37 	&	21.15 (0.38)	&	19.94 (0.42)	&	20.37 (0.47)	&	19.68 (0.31)	&	-	&	-	&	LCO	\\
58674	&	38 	&	-	&	-	&	20.08 (0.03)	&	19.44 (0.03)	&	19.42 (0.14)	&	19.35 (0.14)	&	GROND	\\
58675	&	39 	&	-	&	-	&	-	&	19.67 (0.31)	&	19.86 (0.30)	&	19.89 (0.37)	&	LT	\\
58680	&	44 	&	21.48 (0.73)	&	20.43 (0.66)	&	20.95 (0.71)	&	19.90 (0.45)	&	19.97 (0.46)	&	-	&	LCO	\\
58684	&	48 	&	-	&	20.56 (0.62)	&	21.38 (0.81)	&	20.05 (0.41)	&	20.46 (0.50)	&	-	&	LCO	\\
58685	&	49 	&	-	&	-	&	-	&	20.10 (0.46)	&	20.62 (0.45)	&	20.42 (0.51)	&	LT	\\
58690	&	54 	&	22.31 (0.59)	&	20.93 (0.66)	&	22.13 (1.06)	&	-	&	-	&	-	&	LCO	\\
58694	&	58 	&	-	&	-	&	21.94 (1.00)	&	20.95 (0.49)	&	22.04 (0.71)	&	21.29 (0.65)	&	LT	\\
58697	&	61 	&	22.51 (0.60)	&	-	&	22.91 (1.50)	&	21.37 (0.63)	&	21.55 (0.73)	&	-	&	LCO	\\
58704	&	68 	&	-	&	-	&	-	&	21.57 (0.68)	&	-	&	22.41 (1.09)	&	LT	\\
58713	&	77 	&	-	&	-	&	-	&	21.28 (0.69)	&	21.40 (0.71)	&	-	&	LT	\\
58716	&	80 	&	-	&	-	&	-	&	20.91 (0.55)	&	-	&	-	&	LCO	\\
58725	&	89 	&	23.86 (1.15)	&	22.07 (1.09)	&	-	&	21.52 (0.67)	&	22.59 (1.16)	&	-	&	LCO	\\
58732	&	96 	&	-	&	-	&	-	&	-	&	22.71 (1.28)	&	-	&	LCO	\\
58767	&	131 	&	-	&	-	&	-	&	-	&	-	&	21.76 (0.80)	&	LT	\\
58772	&	136 	&	-	&	-	&	23.03 (1.74)	&	-	&	-	&	22.61 (1.17)	&	LT	\\
\hline
\end{tabular}
\end{table*}

\begin{table*}
\caption{\textit{Swift} AB magnitudes as seen in Figure~\ref{fig: Photometry}.}
\begin{tabular}{c|c|c|c|c|c}
\hline
MJD	&	Phase from maximum (days)	&	UVM2	&	UVW1	&	UVW2	&	u	\\
\hline
58645	&	9 	&	20.72 (0.14)	&	20.21 (0.24)	&	20.71 (0.20)	&	19.68 (0.21)	\\
58651	&	15 	&	20.78 (0.31)	&	>20.39	&	20.72 (0.31)	&	>19.72	\\
58658	&	22 	&	20.80 (0.15)	&	20.50 (0.20)	&	21.17 (0.21)	&	20.54 (0.32)	\\
58660	&	24 	&	21.30 (0.23)	&	20.67 (0.24)	&	21.72 (0.32)	&	>20.58	\\
58663	&	27 	&	20.76 (0.16)	&	20.79 (0.24)	&	20.95 (0.19)	&	>20.70	\\
58666	&	30 	&	21.32 (0.21)	&	20.67 (0.22)	&	21.14 (0.21)	&	>20.71	\\
\hline

\end{tabular}
\end{table*}

\begin{table*}
\caption{ATLAS AB magnitudes as reported in Figure~\ref{fig: Photometry}.}
\begin{tabular}{c|c|c|c}
\hline
MJD	&	Phase from maximum (days)	&	cyan	&	orange	\\
\hline

58609	&	-27 	&	>20.61	&	-	\\
58609	&	-27 	&	>20.69	&	-	\\
58609	&	-27 	&	>20.43	&	-	\\
58609	&	-27 	&	>19.91	&	-	\\
58611	&	-25 	&	-	&	>20.15	\\
58611	&	-25 	&	-	&	>20.16	\\
58611	&	-25 	&	-	&	>19.81	\\
58617	&	-19 	&	-	&	>20.06	\\
58617	&	-19 	&	-	&	>20.12	\\
58617	&	-19 	&	-	&	>20.20	\\
58617	&	-19 	&	-	&	>20.44	\\
58619	&	-17 	&	-	&	>19.71	\\
58619	&	-17 	&	-	&	>19.65	\\
58619	&	-17 	&	-	&	>19.69	\\
58619	&	-17 	&	-	&	>19.74	\\
58620	&	-16 	&	-	&	>19.01	\\
58620	&	-16 	&	-	&	>19.00	\\
58620	&	-16 	&	-	&	>19.13	\\
58620	&	-16 	&	-	&	>19.22	\\
58620	&	-16 	&	-	&	>19.40	\\
58621	&	-15 	&	-	&	>19.41	\\
58621	&	-15 	&	-	&	>19.51	\\
58621	&	-15 	&	-	&	>19.56	\\
58621	&	-15 	&	-	&	>19.60	\\
58623	&	-13 	&	-	&	>17.86	\\
58623	&	-13 	&	-	&	>19.14	\\
58623	&	-13 	&	-	&	>18.98	\\
58623	&	-13 	&	-	&	>19.20	\\
58631	&	-5 	&	-	&	19.25 (0.22)	\\
58631	&	-5 	&	-	&	18.97 (0.24)	\\
58631	&	-5 	&	-	&	18.55 (0.21)	\\
58633	&	-3 	&	18.73 (0.11)	&	-	\\
58633	&	-3 	&	18.83 (0.11)	&	-	\\
58633	&	-3 	&	18.73 (0.11)	&	-	\\
58633	&	-3 	&	18.70 (0.16)	&	-	\\
58637	&	1 	&	18.57 (0.09)	&	-	\\
58637	&	1 	&	18.84 (0.10)	&	-	\\
58637	&	1 	&	18.55 (0.08)	&	-	\\
58637	&	1 	&	18.54 (0.09)	&	-	\\
58643	&	7 	&	-	&	18.90 (0.15)	\\
58643	&	7 	&	-	&	18.87 (0.15)	\\
58643	&	7 	&	-	&	19.11 (0.20)	\\
58643	&	7 	&	-	&	18.96 (0.17)	\\
58645	&	9 	&	-	&	18.80 (0.13)	\\
58645	&	9 	&	-	&	18.69 (0.11)	\\
58645	&	9 	&	-	&	19.14 (0.18)	\\
58645	&	9 	&	-	&	18.79 (0.17)	\\
58647	&	11 	&	-	&	19.16 (0.21)	\\
58647	&	11 	&	-	&	19.16 (0.21)	\\
58649	&	13 	&	-	&	18.90 (0.31)	\\
58649	&	13 	&	-	&	18.81 (0.26)	\\
58659	&	23 	&	-	&	19.01 (0.21)	\\
58659	&	23 	&	-	&	19.54 (0.33)	\\
58659	&	23 	&	-	&	18.96 (0.19)	\\
58659	&	23 	&	-	&	19.33 (0.30)	\\
58659	&	23 	&	-	&	19.05 (0.23)	\\
58659	&	23 	&	-	&	19.53 (0.35)	\\
58665	&	29 	&	20.07 (0.29)	&	-	\\
58665	&	29 	&	19.91 (0.26)	&	-	\\
58665	&	29 	&	19.57 (0.21)	&	-	\\

\hline
\end{tabular}
\end{table*}

\begin{table*}
\begin{tabular}{c|c|c|c}
\hline
MJD	&	Phase from maximum (days)	&	cyan	&	orange	\\
\hline

58667	&	31 	&	-	&	19.62 (0.27)	\\
58667	&	31 	&	-	&	19.70 (0.27)	\\
58667	&	31 	&	-	&	19.17 (0.16)	\\
58667	&	31 	&	-	&	19.31 (0.21)	\\
58669	&	33 	&	19.89 (0.31)	&	-	\\
58669	&	33 	&	19.55 (0.24)	&	-	\\
58669	&	33 	&	19.17 (0.17)	&	-	\\
58670	&	34 	&	19.99 (0.31)	&	-	\\
58670	&	34 	&	20.16 (0.34)	&	-	\\
58671	&	35 	&	19.92 (0.30)	&	-	\\
58671	&	35 	&	20.01 (0.30)	&	-	\\
58671	&	35 	&	19.51 (0.24)	&	-	\\
58674	&	38 	&	-	&	19.49 (0.23)	\\
58674	&	38 	&	-	&	19.62 (0.23)	\\
58674	&	38 	&	-	&	19.69 (0.29)	\\
58685	&	49 	&	-	&	18.97 (0.30)	\\
58723	&	87 	&	-	&	20.23 (0.33)	\\
\hline
\end{tabular}
\end{table*}

%%%%%%%%%%%%%%%%%%%%%%%%%%%%%%%%%%%%%%%%%%%%%%%%%%

% Don't change these lines
\bsp	% typesetting comment
\label{lastpage}
\end{document}